\documentclass[12pt]{article}
\usepackage{amsmath,amssymb,mathtools,hyperref,lscape,pdflscape,enumitem,bbm,graphicx}
\hypersetup{pdfborder = {0 0 0},colorlinks=true,linkcolor=blue,urlcolor=blue,citecolor=blue}

\usepackage[margin=1in]{geometry}
\usepackage[doublespacing]{setspace}
\usepackage[round]{natbib}

\usepackage[justification=centering]{caption}
\usepackage{dcolumn}
\usepackage{multirow}
\usepackage[capposition=top]{floatrow}
\usepackage{float}
\usepackage[caption = false]{subfig}
\usepackage{booktabs}
\usepackage{lscape}
\usepackage{multicol}

\newtheorem{assumption}{Assumption}

\renewcommand{\P}{\mathbbm{P}}

\renewcommand{\a}{\mathsf{a}}
\newcommand{\I}{\mathbbm{1}}

\newcommand{\inputs}{inputs_new}

\begin{document}

\singlespacing

\title{
    \vspace{-0.5in}Randomization Inference for Before-and-After Studies with Multiple Units: An Application to a Criminal Procedure Reform in Uruguay\thanks{A preliminary version of this paper circulated under the title ``Breaking the code: Can a new penal procedure affect public safety.'' We thank Giorgio Chiovelli, Scott Cunningham, Jennifer Doleac, Libor Du\v{s}ek, Jeff Grogger, Dorothy Kronick, Emily Owens, Zachary Peskowitz, Yotam Shem-Tov, Gonzalo Vazquez-Bare and participants at the $2021$ Chicago/LSE Conference on the Economics of Crime and Justice, the $2021$ Annual Meeting of the Latin American and Caribbean Economic Association, the $2021$ LACEA/RIDGE Workshop on the Economics of Crime, the $2021$ APSA Annual Meeting, the $2020$ Annual Conference of the Latin American Society for Political Methodology, the Summer $2020$ Online Seminar on the Economics of Crime, and Syracuse University, Universidad de Chile, Universidad ORT, and Universidad de Montevideo Economics Department Seminar Series for thoughtful comments and suggestions. The Editor, Associate Editor, and reviewers provided valuable suggestions that help improved the paper. Cattaneo and Titiunik gratefully acknowledge financial support from the National Science Foundation (SES-2241575). Díaz gratefully acknowledges financial support from the National Agency of Research and Development’s (ANID) Millennium Science Initiative Program (NCS2024\_058).}
}
\author{
    Matias D. Cattaneo\thanks{Department of Operations Research and Financial Engineering, Princeton University.}\and
    Carlos D\'iaz\thanks{Millennium Nucleus on Criminal Complexity and School of Economics and Business, Universidad Alberto Hurtado.}\and
    Roc\'io Titiunik\thanks{Department of Politics, Princeton University.}
}

\date{\today }

\maketitle

\begin{abstract}
    Learning about the immediate causal effects of large-scale policy interventions poses a significant challenge for quasi-experimental methods that rely on long-term trends or parametric modeling assumptions. As an alternative, we develop a randomization inference framework for before-and-after studies with multiple units, designed specifically for short-term causal inference and allowing for general assignment mechanisms. The method provides finite-sample-valid statistical inferences without relying on parametric time series models or extrapolation. We demonstrate its utility by analyzing a major criminal justice reform in Uruguay that switched from an inquisitorial to an adversarial system in November $2017$. Our method relies on the key assumption of no local time trends near the policy adoption time, which is supported by several falsification tests in our empirical study. We find a statistically significant short-term causal effect: an increase of approximately $25$ daily police reports (an $8\%$ rise) in the first week of the new justice system. Our randomization inference framework provides a robust and flexible methodology for evaluating policy adoptions in before-and-after studies with multiple units.
\end{abstract}


\thispagestyle{empty}

\clearpage
\tableofcontents
\thispagestyle{empty}

\clearpage
\doublespacing
\setcounter{page}{1}

\section{Introduction} \label{intro}

On November 1st, 2017, a new code of criminal procedure (CCP or \textit{C\'odigo del Proceso Penal} in Spanish) became effective in Uruguay. The reform ended the old inquisitive and written tradition, and established an accusatory, adversarial, oral, and public criminal system in its place. Under the old CCP, the inquisitorial judge led the investigation and decided the appropriate punishment for a particular crime. Once the new criminal procedure came into effect, judges became neutral referees focused on ensuring the correct procedure, while prosecutors became responsible for leading the investigation. Under the new regime, the investigation is the exclusive responsibility of prosecutors who, representing society, must present evidence to judges; judges then decide what evidence to admit into the record. This separation between the roles of prosecutor and judge is a key reason why adversarial systems are generally considered fairer and less susceptible to abuse than inquisitorial systems.

Despite its advantages, this type of reform can lead to unintended changes in the costs associated with offending. The new CCP may have changed both the severity and the certainty of punishments through several channels, possibly affecting the propensity and ability of individuals to commit crimes. For instance, the new adversarial system introduced substantial changes to the adjudication process of criminal law such as plea bargain, alternatives to oral trials, and exceptional use of preventive detention, all of which might result in lighter sentences. In addition, prosecutors faced both a new role and a significant increase in their workload, while policing was affected by new rules and supervision of police investigations. These changes resulted in coordination problems between prosecutors and police officers during the first months of the adversarial system, possibly affecting the probability of detection and conviction. 

We propose a methodology to evaluate the immediate impact of this procedural reform on the number of offenses reported to the police in Montevideo, the capital and largest city of Uruguay. The reform was introduced nationwide simultaneously but, in contrast to a standard interrupted time series design \citep[e.g.][]{ShadishCookCampbell2002-book}, we observe multiple units in each time period because we collect crime reports (\textit{denuncias}) for each of the $62$ neighborhoods in Montevideo. This leads to a before-and-after observational study with multiple units: a setting where several cross-sectional units are first observed for several periods, a policy is then adopted at the same time for all units, and finally, the same units are observed after the intervention for several more periods. 

This type of observational study is often analyzed using linear panel data models where the outcome for unit $i$ in period $t$ is typically regressed on a unit's fixed effect, a time fixed effect, time trends, unit-level covariates, and time-indexed treatment indicators---see \cite{FreyaldenhovenEtal2019-AER,FreyaldenhovenEtal2024-wp} and \cite{Miller2023-JEP} for reviews. This empirical strategy requires many units and many time periods, and relies on time homogeneity over a typically long time span. The number of parameters to be estimated is often large, and multicollinearity issues are common. A relatively small number of units can make estimation of the time fixed effects unstable, while a similar problem occurs for the estimation of unit fixed effects with few time periods. The linear panel event study approach also relies on a parametric specification to model the global trend of the outcome to separate it from the effect of the policy. As a consequence, this observational method is most useful when many units are observed over many time periods, and the time homogeneity of the flexible parametric regression model is plausible, allowing researchers to model the time series globally over a long time span.  Furthermore, a crucial limitation of before-and-after studies is the lack of a control group \citep{Miller2023-JEP}, which suggests focusing on small windows of time around the intervention to isolate the treatment effect from potential time confounders. 

We consider an alternative approach that is tailored to learning about the immediate effect of the new CCP on the number of police reports, while allowing for a small number of units and time periods, avoiding global time series specifications, and providing robustness to cross-sectional and time-series dependence. Building on the causal inference literature \citep{Rosenbaum_2002_book,Rosenbaum2010-book}, we develop a Fisherian randomization inference approach for before-and-after studies with multiple units. Our approach is based on localizing around the time of the policy adoption rather than modeling the time series globally, which serves two purposes simultaneously: (i) explicitly targets the immediate causal effect of the policy, and (ii) removes the need for parametric time series modeling. In our framework, the units' potential outcomes are non-stochastic, and we assume that the policy adoption time can be approximated by a known distribution.

While in reality all units are untreated in the pre-intervention period and treated in the post-intervention period, we consider two hypothetical assignment mechanisms: in the \textit{Treatment Reversal} (TR) assignment mechanism, each unit could have been treated either in the pre-intervention or in the post-intervention period, while in the \textit{Adoption Timing} (AT) assignment mechanism, each unit is always untreated before adoption and treated afterwards, but the timing of the intervention could have been sooner or later than the actual adoption time. Because the potential outcomes are non-stochastic, both assignment mechanisms provide the distribution of any test statistic under suitable assumptions and null hypotheses, providing valid p-values to hypothesis testing. Moreover, our framework is general and allows for other assignment mechanisms in addition to the two we propose.

Our hypothesized randomization mechanisms are more easily justified for a small window around the time of adoption of the new policy. Moreover, the TR mechanism specifically relies on the assumption that the average potential outcomes in a window around adoption time exhibit no trends, which is implausible when the window is large. Because we observe multiple units for every time period, we can analyze the smallest possible window around the time of the intervention, thereby reducing extrapolation to a minimum. To guide the choice of windows larger than the minimum, we propose a procedure based on the procedure by \cite{Cattaneo-Frandsen-Titiunik_2015_JCI}: choosing an artificial (or `placebo') adoption time well before the adoption of the real intervention, we test the null hypothesis of no treatment effect for the main outcome for a sequence of nested windows of increasing length, stopping when the hypothesis is rejected. Because the treatment effect is known to be zero, this procedure guides the choice of window around the real adoption time under the assumption that time trends in the placebo windows are the same as in our real window. We show that, for various artificial adoption times, windows of length up to approximately $14$ days within the cutoff have outcome trends that cannot generally be distinguished from zero. We further assess our assumptions with a falsification analysis that, using the chosen window length, replicates the analysis using data from the past.

We find that the implementation of Uruguay's new criminal procedure resulted in a local increase in the number of crimes reported to the police in Montevideo of about $8.2$ percent in the week after the reform. These estimates are consistent with the view that legal codes are far from innocuous \citep{acemoglu2005unbundling}. Our randomization-based approach indicates that the effect of the policy is distinguishable from zero within seven days of the intervention with the TR mechanism (p-value $\leq 0.02$) and within fourteen days with both the TR and the AT mechanism (p-values $\leq 0.03$). Our inferences also reject the null hypothesis of no effect when multiple windows within fourteen days are considered jointly.

Our focus on immediate or short-run effects is well suited to before-and-after designs, which are typically employed to study wide-ranging policy reforms that are adopted everywhere in a country or region. The immediate effects of such policies are often critical for the reform's trajectory, as they have the potential to shape public opinion and influence its long-term success o failure \citep{jain2005public, stokes1996public, stokes2001public}. In our case study, the documented increase in crime reports appears to have affected the perceived legitimacy of the new adversarial system during its initial phase, which led to future changes and impacted the ultimate features of the new adversarial regime. More generally, when a major legal reform is implemented during a time of heightened public concern, short-term results may have long term consequences. Unrealistic expectations can quickly turn into frustration when early indicators fail to show progress, raising the risk of backsliding or even reversal.

\subsection{Related Literature}\label{sec:Related Literature}

Methodologically, our approach contributes to a rich literature on randomization inference methods for program evaluation and causal inference, including methods for the analysis of natural experiments \citep{Ho-Imai_2006_JASA}, instrumental variables \citep*{ImbensRosenbaum_2005_jrssA,kang2018inference}, and regression discontinuity designs \citep{Cattaneo-Frandsen-Titiunik_2015_JCI,Cattaneo-Titiunik-VazquezBare_2017_JPAM}. Our approach is also connected to the so-called regression discontinuity (RD) design in time \citep{HausmanRapson2018-AnnRevResEcon}, where there are multiple cross-sectional units and the running variable is defined as time to the event---a setting that is not a standard RD design \citep{Cattaneo-Titiunik_2022_ARE}. \citet{HausmanRapson2018-AnnRevResEcon} identify a dilemma in applying standard RD methods to before-and-after studies: using a large region of data around the policy change can lead to extrapolation, while using a local region often leaves too few observations for valid asymptotic approximations. Our method circumvents these limitations by localizing near the treatment time, avoiding the extrapolation inherent in standard parametric approaches. See \citet[Section 2]{Cattaneo-Idrobo-Titiunik_2024_Book} for an introduction to local randomization methods in the context of standard RD designs.

Substantively, our research adds to the economics of crime literature by showing how procedural law impacts delinquency. More lenient punishments---often associated with plea bargaining \citep{ bushway2012plea}---and a lower probability of being caught and convicted have been argued to increased crime \citep{Becker_1968}. Prior empirical studies have shown that changes in evidentiary rules, trial length, court efficiency, and enforcement incentives can all affect criminal behavior \citep{atkins2003effects, Dalla, Dusek, Soares}. In Latin America, the shift to adversarial systems has produced mixed results \citep{Langer}. For example,  \citep{Zorro} documents increases in crime due to reduced detection rates in Colombia, while \cite{Kronick} finds no effects on crime despite higher arrest rates.

\section{Criminal Procedure Reform in Uruguay} \label{background}

Uruguay was one of the last countries in Latin America to reform its code of criminal procedure. Beginning in the early 1990s, procedural reforms spread across the region and most countries transitioned from inquisitorial to adversarial systems---with the notable exceptions of Brazil and Cuba, which still preserve their traditional frameworks \citep{fandino2020adversarial}. This shift constituted a major transformation of the region's criminal justice system \citep{Langer}.

Historically, Latin America's criminal procedures were governed by inquisitorial, written systems introduced in the nineteenth and early twentieth centuries. As described by \citet{Langer}, these systems typically shared two core features, both of which also characterized the Uruguayan CCP. First, proceedings were divided into two written phases: a pretrial investigation phase and a verdict-and-sentencing phase, with the latter largely based on a dossier compiled by judges and police. Second, the judge played a dual role---leading the investigation and later adjudicating the case. The defendant was excluded from the investigation, which remained secret, and pretrial detention was the norm. Over time, however, the rising prominence of human rights (in the 1970s) and the region's democratization processes (in the 1980s and 1990s) led to mounting concerns that such systems failed to uphold basic due process guarantees \citep{maier2000introduccion, Langer}.

In response, countries began to adopt adversarial, accusatory, oral, and public systems, featuring three well-defined stages: the formalization of the investigation, a preliminary hearing, and a trial phase \citep{fandino2020adversarial}. Under this model, the preliminary investigation is conducted by the police and led by a prosecutor, and the defense and prosecution present their arguments in public hearings. Uruguay’s reformed system closely follows this structure. During the first stage, the General Prosecution Office decides whether to pursue criminal charges (i.e., the judicialization of the case). If the investigation is formalized, the second stage consists of a preliminary hearing to review investigative outcomes. Finally, an oral trial is held for cases not resolved through alternative outcomes. Uruguay has a classic adversarial procedure where defense and prosecution arguments are presented at a hearing.

The adoption of Uruguay's adversarial CCP marked a complex and far-reaching institutional reform. Although the law was passed on December 19, 2014, the new system did not take effect until November 1, 2017. The scale and scope of the changes required nearly three years of preparation. The reform redefined the roles of key actors in the criminal justice system---particularly prosecutors and police officers. Under the new criminal adjudication law, prosecutors became responsible for directing investigations, representing a fundamental change in both their function and workload. The transition was so demanding that the prosecutors' union publicly opposed the reform, citing overwhelming workloads. Some prosecutors were assigned several hundred cases simultaneously, leading many to take mental health leave during the first year of implementation \citep{solomita}. Police officers, for their part, were required to adapt to new rules and supervision, with a prosecutor overseeing the case instead of a judge. In sum, the reform affected many aspects of the criminal system. For this reason, when we refer to the effect of the CCP reform, we mean the combined effect of the bundle of treatments that were changed as part of the reform, not simply the adoption of a more due-process-oriented system. 

Important for the interpretation of our results, the public was intentionally informed about the reform, as the government engaged in efforts to disseminate information and bring public awareness. The Office of the Attorney General launched a public information campaign, as a result of which the reform received considerable media coverage. In Section SA-4 in the Supplemental Appendix, we show a billboard that was placed in Montevideo bus stops highlighting the transparency and guarantees of the new procedural law.

Uruguay experienced an unprecedented increase in the number of police reports following the implementation of the new CCP. While there was broad consensus around the normative desirability of transitioning to an adversarial model, the reform was blamed by public officials for the observed rise in insecurity. In particular, both the President and the Interior Minister referred to the sharp increase in police reports as the ``November Effect'', invoking the date of adoption of the reform \citep{Palumbo}. Although the short-term impacts led to eventual modifications of the policy such as clarifying guidelines from the Attorney General and even legislative amendments, these changes occurred well outside the time periods we analyze. Our approach does rely on the assumption that the nature of the reform stayed constant in the window around the reform's adoption time that we use for analysis; we are not aware of any modifications to the implementation of the reform during this period.

\subsection{Data and Overall Patterns} \label{data}

We collected data on the offenses reported to the police in Montevideo from the Ministry of Interior of Uruguay. Montevideo is the capital and largest city of Uruguay, home to 40\% of Uruguay's total population. We first present a descriptive analysis of the years before and after the adoption of the reform. Since the new CCP came into effect on November 1st, 2017, we start by building a two-year symmetric window from November 1st, 2016 trough October 31st, 2018. Table~\ref{T.01} reports the average number of crimes reported daily to the police during this overall period, as well as for the years before and after the switch to the new CCP. An average of 331 crimes were reported to police every day during this two-year period, with three categories accounting for more than 7 of every 10 reports processed in Montevideo: theft ($\approx46\%$ of all offenses reported to police), robbery ($\approx16\%$) and domestic violence ($\approx 10\%$).

\vspace{10pt}
\begin{table}[!htbp]
\captionsetup{justification=centering}
\caption{Average Daily Crimes Reported to Police in Montevideo}
\def\sym#1{\ifmmode^{#1}\else\(^{#1}\)\fi}
\label{T.01}\centering
\par
{\footnotesize
\begin{tabular}{lcccc}
\toprule
 	   &\multicolumn{1}{c}{Two-Year Window}&\multicolumn{1}{c}{Old CCP}&\multicolumn{1}{c}{New CCP}&\multicolumn{1}{l}{$\quad\,\Delta$}\\
 	   &\multicolumn{1}{c}{\scriptsize 11/01/16 to 10/31/18}&\multicolumn{1}{c}{\scriptsize(a)}&\multicolumn{1}{c}{\scriptsize(b)}&\multicolumn{1}{l}{\scriptsize \, (b)-(a)}\\
\cmidrule(l){2-2} \cmidrule(l){3-5}
{\it Theft } &\multicolumn{1}{c}{152.19}&\multicolumn{1}{c}{130.05}&\multicolumn{1}{c}{174.33}&\multicolumn{1}{c}{44.28}\\
	   &\multicolumn{1}{c}{\scriptsize \it 46\%}&\multicolumn{1}{c}{\scriptsize (0.942)}&\multicolumn{1}{c}{\scriptsize (1.221)}&\multicolumn{1}{l}{\scriptsize (1.543)}\\   
\addlinespace
{\it Robbery } &\multicolumn{1}{c}{51.81}&\multicolumn{1}{c}{41.60}&\multicolumn{1}{c}{62.02}&\multicolumn{1}{c}{20.42}\\
	   &\multicolumn{1}{c}{\scriptsize \it 16\%}&\multicolumn{1}{c}{\scriptsize (0.426)}&\multicolumn{1}{c}{\scriptsize (0.614)}&\multicolumn{1}{l}{\scriptsize (0.747)}\\
\addlinespace
{\it Domestic Violence } &\multicolumn{1}{c}{33.66}&\multicolumn{1}{c}{32.79}&\multicolumn{1}{c}{34.53}&\multicolumn{1}{c}{1.75}\\
	   &\multicolumn{1}{c}{ \scriptsize \it 10\%}&\multicolumn{1}{c}{\scriptsize (0.379)}&\multicolumn{1}{c}{\scriptsize (0.382)}&\multicolumn{1}{l}{\scriptsize (0.539)}\\
\addlinespace
{\it Other Crimes}  &\multicolumn{1}{c}{92.98}&\multicolumn{1}{c}{78.98}&\multicolumn{1}{c}{106.99}&\multicolumn{1}{c}{28.02}\\
	   &\multicolumn{1}{c}{\scriptsize \it 28\%}&\multicolumn{1}{c}{\scriptsize (0.662)}&\multicolumn{1}{c}{\scriptsize (0.869)}&\multicolumn{1}{l}{\scriptsize (1.092)}\\
\addlinespace
\cmidrule(lr){1-1}\cmidrule(lr){2-2} \cmidrule(lr){3-5}
Total Police Reports &\multicolumn{1}{c}{ 330.63}&\multicolumn{1}{c}{283.41}&\multicolumn{1}{c}{377.87}&\multicolumn{1}{c}{94.47}\\
&\multicolumn{1}{c}{\scriptsize \it 100\%}&\multicolumn{1}{c}{\scriptsize (1.524)}&\multicolumn{1}{c}{\scriptsize (2.013)}&\multicolumn{1}{l}{\scriptsize (2.525)}\\
\addlinespace
Days&\multicolumn{1}{c}{730}&\multicolumn{1}{c}{365}&\multicolumn{1}{c}{365}&\\
\cmidrule(lr){1-5}
\multicolumn{5}{l}{\scriptsize Percentage of total number of reports in {\it italic}; standard errors in parentheses.}\\
\multicolumn{5}{l}{\scriptsize Old CCP: inquisitorial system (November 1st, 2016 to October 31st, 2017).}\\ 
\multicolumn{5}{l}{\scriptsize New CCP: adversarial system (November 1st, 2017 to October 31st, 2018).}\\
\multicolumn{5}{l}{\scriptsize Source: Ministry of Interior of Uruguay.}
\end{tabular}
}
\end{table}

Table~\ref{T.01} shows a considerable increase in the number of police reports in Montevideo during the first year of the new CCP. When we compare the last year of the old system to the first year of the new system, the total number of daily reported offenses increases by approximately 95 incidents (from roughly 283 to 378 per day), a statistically significant difference that represents a 33\% increase. This upward trend is also reflected in the three most frequent crimes reported in Montevideo. Regarding property crimes, the number of thefts and robberies reported to police every day increased by 34\% and 49\%, respectively. Reports of domestic violence, the most frequent crime against the person, exhibited a much smaller increase of 5\%.

An initial visual inspection of the data suggests a strong association between the increase in police reports and the timing of the CCP reform. Figure~\ref{F.01} plots the daily number of offenses reported to the police in Montevideo from 600 days before the new CCP came into effect (early 2016) until 400 days after (late 2018).

\vspace{5pt}
\begin{figure}[!ht]
    \begin{center}
        \caption[Caption for LOF]{Crime in Montevideo\\ {\small (total police reports, 2016-2018)}}	  
        \vspace{-10pt}
        \includegraphics[width=1\textwidth]{\inputs/CDT_Figure_TotalReports1618.png}
        \label{F.01}
    \end{center}
\end{figure}
Figure~\ref{F.01} suggests an increase in the number of police reports shortly after the new CCP came into effect, consistent with the previously mentioned ``November Effect.'' To identify a potential immediate causal effect, all our analyses will focus on time windows contained within 21 days before and 21 days after the date of implementation of the new CCP, as marked by the dashed red lines in Figure~\ref{F.01}.

\section{Randomization Inference Framework} \label{sec:methods}

Given the possibility of time trends and other confounders, the patterns reported in the prior section cannot be taken as conclusive evidence that Montevideo's crime wave was caused by the changes to Uruguay's CCP. We develop a randomization inference framework to provide further empirical evidence that may address these (and other) methodological concerns. Our framework is designed to analyze a before-and-after study with multiple units, that is, a setting where multiple cross-sectional units are first observed for some periods, a reform affecting all units is introduced at the same time for all units, and then the same units are observed after the reform for some more periods.

We adopt a Fisherian framework with non-random potential outcomes. Each cross-sectional unit $i=1,2,\ldots, n$ is observed over $t = 1, 2, \ldots, T$ time periods, and an intervention is adopted at time period $t=\a_0 \in\{1,2,\ldots,T\}$.  The fixed treated and untreated potential outcomes at time $t$ for unit $i$ are denoted, respectively, by $y_{i,t}(1)$ and $y_{i,t}(0)$. This notation implicitly assumes that the specific time of adoption $\a_0$ can only affect the potential outcomes indirectly via the triggering of the intervention, but not directly---an exclusion restriction that is most plausible within small time windows around $\a_0$. In our sample, $n=62$ neighborhoods, and the new CCP began precisely at 12:00 AM on November 1st, $2017$, becoming active at the same specific moment for all units; thus, in the actual policy assignment, all units are untreated in all periods $t<\a_0$ and treated in all periods $t\geq \a_0$.

Assuming enough time periods are available, we define a window including $\tau$ periods before and $\tau$ periods after the time of policy adoption,
\begin{align*}
    \mathcal{W}_\tau = \{\a_0-\tau, \ldots, \a_0-1,\a_0, \a_0+1,\ldots, \a_0+\tau-1\},
\end{align*}
where we consider symmetric windows only for simplicity. The length of $\mathcal{W}_\tau$ will depend on the unit of time. Since each unit appears in every period, the total sample size for $\mathcal{W}_\tau$ is $2\tau n$ observations. For example, in our application $t$ measures days, $\mathcal{W}_1 = \{\a_0-1,\a_0 \}$ includes the day immediately before adoption and the adoption day, and hence there are $124$ observations within $\mathcal{W}_1$.

We denote the treatment indicator by $D_{i,t}$, which is equal to one if unit $i$ is treated at time $t$ and zero otherwise. This is the random variable that determines the assignment of the intervention in each period for each unit. The (random) observed outcome is
\begin{align*}
    Y_{i,t} =  D_{i,t} \cdot y_{i,t}(1)  + (1-D_{i,t}) \cdot y_{i,t}(0).
\end{align*}
In a before-and-after design, all units are untreated during some periods and treated during others. For each unit, we define two average observed outcomes within $\mathcal{W}_\tau$:
\begin{align*}
    \bar{Y}_{i,\tau,1} = \frac{1}{N_{i,\tau,1}} \sum_{t \in \mathcal{W}_\tau} D_{i,t} Y_{i,t},
    \qquad
    N_{i,\tau,1} = \sum_{t \in \mathcal{W}_\tau} D_{i,t},
\end{align*}
and
\begin{align*}
    \bar{Y}_{i,\tau,0} = \frac{1}{N_{i,\tau,0}} \sum_{t \in \mathcal{W}_\tau} (1-D_{i,t})Y_{i,t},
    \qquad
    N_{i,\tau,0} = \sum_{t \in \mathcal{W}_\tau} (1-D_{i,t}),
\end{align*}
which correspond to the average outcome across all periods $t \in \mathcal{W}_\tau$ during which the unit is treated, and the average outcome across all periods $t \in \mathcal{W}_\tau$  during which the unit is untreated, respectively. 

According to the actual intervention, $D_{i,t} = \I(t \geq \a_0)$ for all $i=1,2,\ldots, n$. Our approach involves hypothesizing different possible assignment mechanisms governing $D_{it}$. For a time of adoption $a$, we define four average potential outcomes within the window $\mathcal{W}_\tau$ for every unit, corresponding to treated and untreated potential outcomes averaged over the pre-intervention ($t < a$) or post-intervention ($t\geq a$) periods:
\begin{align*}
    \bar{y}_{i,\tau, -}(0,a)
    = \frac{1}{n_{-}(a)} \sum_{t \in \mathcal{W}_\tau} \I(t< a) y_{i,t}(0),
    \qquad
    \bar{y}_{i,\tau, -}(1,a)
    = \frac{1}{n_{-}(a)} \sum_{t \in \mathcal{W}_\tau} \I(t< a) y_{i,t}(1),
\end{align*}
with $n_{-}(a) = \sum_{t \in \mathcal{W}_\tau} \I(t<a)$, and
\begin{align*}
    \bar{y}_{i,\tau, +}(0,a)
    = \frac{1}{n_{+}(a)} \sum_{t \in \mathcal{W}_\tau} \I(t\geq a) y_{i,t}(0),
    \qquad
    \bar{y}_{i,\tau, +}(1,a)
    = \frac{1}{n_{+}(a)} \sum_{t \in \mathcal{W}_\tau} \I(t\geq a) y_{i,t}(1),
\end{align*}
with $n_{+}(a)=\sum_{t \in \mathcal{W}_\tau} \I(t\geq a)$, and where $n_{-}(a)+n_{+}(a)=2\tau$ for all $a\in \mathcal{W}_\tau$. Of these average potential outcomes, we only observe $\bar{y}_{i,\tau,-}(0,\a_0)$ and $\bar{y}_{i,\tau,+}(1,\a_0)$ for $i=1,\ldots,n$; all the others are counterfactual and thus unobserved.

Our notation explicitly distinguishes between observed and potential averages: the subindeces $0$ and $1$ used in $\bar{Y}_{i,\tau,1}$ and $ \bar{Y}_{i,\tau,0}$ denote, respectively, untreated and treated  observed average outcomes whenever they may occur, while the $-$ and $+$ subindeces in the non-random averages $\bar{y}_{i,\tau, +}(\cdot,a)$ and $\bar{y}_{i,\tau, -}(\cdot,a)$ denote, respectively, pre-intervention and post-intervention periods relative to adoption time $t=a$, regardless of whether the periods entering the averages were untreated or treated according to the assignment $D_{it}$.

Finally, to describe different assignment mechanisms, we define the vector $\mathbf{D}_{t} = (D_{1,t}, D_{2,t}, \ldots,D_{n,t})'$ for each period $t \in \mathcal{W}_\tau$ and collect them into the $n \times 2\tau$ matrix
\begin{align*}
    \mathbf{D}
    = [\mathbf{D}_{\a_0-\tau}, \ldots, \mathbf{D}_{\a_0-1}, \mathbf{D}_{\a_0}, \mathbf{D}_{\a_0+1}, \ldots,\mathbf{D}_{\a_0+\tau-1}].
\end{align*}
In an unrestricted setting where each unit could be assigned to control or treatment in any period, we could observe zeros or ones in any of the entries of $\mathbf{D}$ regardless of the values taken by the other entries, resulting in $2^{2n\tau}$ different possible assignment matrices. A before-and-after design, however, places known restrictions on the columns of $\mathbf{D}$. Our approach to inference considers two different assignment mechanisms that restrict the matrix $\mathbf{D}$ in different ways, and lead to two different randomization distribution of test statistics that can be used to test null hypotheses. Other assignment mechanisms in addition to the ones we describe can be accommodated with different restrictions $\mathbf{D}$. 

\subsection{Treatment Reversal (TR) Assignment Mechanism}

The first assignment mechanism keeps the adoption time at $t=\a_0$ for all units and hypothesizes a possible reversal of the treatment condition between the period before and the period after $\a_0$. For each unit, this mechanism independently assigns the treatment to either all pre-intervention periods or all post-intervention periods in $\mathcal{W}_\tau$. For example, if the window is $\mathcal{W}_2=\{\a_0-2,\a_0-1,\a_0,\a_0+1\}$, the TR mechanism assigns all the post-intervention periods ($t \in \{\a_0,\a_0+1\}$) to treatment---and therefore assigns all the pre-intervention periods ($t \in \{\a_0-2,\a_0-1\}$) to control---with some known probability. Although in the realized assignment all neighborhoods in Montevideo started the new CCP on November 1, 2017, this mechanism imagines that the assignment of neighborhoods to treatment could have been reversed, with some neighborhoods instead assigned to the old CCP on November 1, 2017, and the new CCP before this date.

Every unit thus has two possible assignments, $\{t: t < 0, t \in  \mathcal{W}_\tau\}$ treated and $\{t: t \geq 0, t \in  \mathcal{W}_\tau\}$ control, or vice-versa. To formalize, we define a binary random variable $Z_i$ that is equal to one when unit $i$ receives the treatment and is equal to zero otherwise. The treatment indicator becomes
\begin{align*}
    D_{i,t} = (1-Z_i) \I(t< \a_0) + Z_i \I(t\geq \a_0),
    \qquad
    t \in \mathcal{W}_\tau,
\end{align*}
for all units $i=1,\ldots,n$.

In this assignment mechanism, if the first post-intervention period is treated (untreated), we know all subsequent periods are treated (untreated) and all pre-intervention periods are untreated (treated). Thus, $D_{i,t} = D_{i,\a_0-1}$ if $t < \a_0$, and $D_{i,t} = D_{i,\a_0}$ if $t \geq \a_0$, for $t \in \mathcal{W}_\tau$. It follows that $D_{i,-1} + D_{i,0} =1$, that is, exactly one of the two periods before and after adoption time is assigned to treatment. These restrictions reduce the number of possible assignment matrices $\mathbf{D}$ from $2^{2\tau n}$ to $2^{n}$; we collect these matrices in the set $\mathcal{D}_{\texttt{TR}}$. Because we assume that each matrix assignment is equally likely, we have $\P[ \mathbf{D}  = \mathbf{d} ] = 2^{-n}$ for the allowable $\mathbf{d} \in \mathcal{D}_{\texttt{TR}}$. 

The conditions for using a randomization inference approach under the TR assignment mechanism are summarized in the following assumption.

\begin{assumption}[TR Assignment Mechanism]\label{ass:AM1}
    There exists a window $\mathcal{W}_\tau$ of $\tau$ periods before and after the adoption time $\a_0$ of the intervention such that the following conditions hold for all $i=1,2,\cdots,n$.
    \begin{enumerate}[label=\emph{(\roman*)},leftmargin=*]
        \item $\bar{y}_{i,\tau,-}(0,\a_0)$, $\bar{y}_{i,\tau,-}(1,\a_0)$, $\bar{y}_{i,\tau,+}(0,\a_0)$ and $\bar{y}_{i,\tau,+}(1,\a_0)$ are non-stochastic. 

        \item $\bar{y}_{i,\tau,-}(0,\a_0) = \bar{y}_{i,\tau,+}(0,\a_0)$ and $\bar{y}_{i,\tau,-}(1,\a_0) = \bar{y}_{i,\tau, +}(1,\a_0)$.
        
        \item $\P[ \mathbf{D}  = \mathbf{d} ] = 2^{-n}$ for all $\mathbf{d} \in \mathcal{D}_{\texttt{TR}}$.
    \end{enumerate}
\end{assumption}

Assumption \ref{ass:AM1}(i) allows for the deployment of Fisherian inference methods because it fixes the potential outcomes. Assumption \ref{ass:AM1}(ii) rules out time trends in the average potential outcomes, which allows us to distinguish the effect of the treatment from over-time changes in the outcomes and thus impute all missing potential outcomes under our maintained null hypothesis. This assumption will be most plausible for relatively small windows $\mathcal{W}_\tau$. The choice of the window $\mathcal{W}_\tau$ is therefore crucial; we discuss it in Section \ref{subsec:methods_Wsel}. Finally, Assumption \ref{ass:AM1}(iii) sets the assignment mechanism to the TR mechanism. 

We consider the null hypothesis that the average treated potential outcome is equal to the average untreated potential outcome in the $\tau$-length post-intervention period for every unit. Formally,
\begin{align*}
    \mathsf{H}_{\mathtt{TR}}: \bar{y}_{i,\tau,+}(0,\a_0) = \bar{y}_{i,\tau,+}(1,\a_0) \qquad\text{for all $i=1,2,\ldots, n$}.
\end{align*}
Under Assumption \ref{ass:AM1}, this hypothesis is sharp, and the average potential outcomes can be imputed for all realizations of the matrix $\mathbf{D}$. It follows that for any test statistic based on the observed average outcomes,  $\mathbf{D}$ will be the only source of randomness.

Following standard randomization inference ideas, we define a test statistic $S(\mathbf{D},\bar{\mathbf{Y}}_{0},\bar{\mathbf{Y}}_{1})$, where $\bar{\mathbf{Y}}_s = (\bar{Y}_{1,\tau,s},\ldots,\bar{Y}_{n,\tau,s})$ for $s=0,1$ are the vectors that collect the observed average untreated and treated outcomes. Under Assumption \ref{ass:AM1} and $\mathsf{H}_{\mathtt{TR}}$, $S(\mathbf{D},\bar{\mathbf{Y}}_{0},\bar{\mathbf{Y}}_{1}) = S(\mathbf{D},\bar{\mathbf{y}}(\a_0))$, where $\bar{\mathbf{y}}(\a_0) = (\bar{y}_{1,\tau,-}(0,\a_0), \ldots, \bar{y}_{n,\tau,-}(0,\a_0),\bar{y}_{1,\tau,+}(1,\a_0), \ldots, \bar{y}_{n,\tau,+}(1,\a_0))'$, and therefore the distribution of $S(\mathbf{D},\bar{\mathbf{Y}}_{0},\bar{\mathbf{Y}}_{1})$ is fully determined by the randomization distribution of $\mathbf{D}$. The test statistic $S$ is a function of the average observed outcomes because Assumption \ref{ass:AM1} places restrictions on the average potential outcomes, which is a simple way to think about (local to $t=\a_0$) time trends. However, one could change the assumption to make it on the individual (non-average) outcomes, in which case the test statistic could be more general.

We denote the observed, realized value of the matrix $\mathbf{D}$ as $\mathbf{d}_\mathtt{obs}$, and the observed value of test statistic as $s_\mathtt{obs} = S(\mathbf{d}_\mathtt{obs},\bar{\mathbf{Y}}_0,\bar{\mathbf{Y}}_1)$. 
The exact two-sided p-value is given by
\begin{align*}
    p_\mathtt{TR}
    = \P\big[|S(\mathbf{D},\bar{\mathbf{Y}}_0,\bar{\mathbf{Y}}_1)| \geq |s_\mathtt{obs}| \big]
    = \frac{1}{2^n} \sum_{\mathbf{d}\in\mathcal{D}_{TR}} \I(|S(\mathbf{d},\bar{\mathbf{y}}(\a_0))| \geq |s_\mathtt{obs}|),
\end{align*}
which can be calculated for every realization of $\mathbf{D}$ based on observed data. For even moderate values of $n$, the total number of possible treatment assignment vectors is too large, so complete enumeration is typically unfeasible. In our application, $n=62$, and hence we compute the p-value by simulation. In each simulation, we construct one of the possible treatment assignment vectors using the postulated assignment mechanism, and computing the value of the test statistic. We repeat this simulation 10,000 times, and then calculate the share of the 10,000 test statistics that have absolute value equal to or higher than $|s_\mathtt{obs}|$.

\subsection{Adoption Timing (AT) Assignment Mechanism}

The TR assignment mechanism allows the treated period to come before the untreated period, a feature that does not respect the chronological ordering of the policy in the real assignment. We consider an alternative mechanism, the Adoption Timing (AT) assignment mechanism, where we retain the time ordering of the event study, and instead hypothesize that the adoption time could have happened sooner or later than $\a_0$ for each unit. The AT assignment mechanism is based on a random time of adoption, which is represented by a discrete uniform random variable $A_i$ with support $\mathcal{A} \subseteq\mathcal{W}_\tau$, allowing for up to $\tau -1$ back-dating periods and up to $\tau$ forward-dating periods from the actual adoption time $t=\a_0$. For example,  $\mathcal{A} = \{-1,0,1\}$ would allow backdating one period, staying at the actual adoption time, or forward-dating one period, while $\mathcal{A} = \{-2,-1\}$ would only allow back-dating up to two periods. Letting $J$ denote the number of periods in $\mathcal{A}$, we have $\P[A_i=a] = \frac{1}{J}$ for all $a \in \mathcal{A}$. Allowing the generality of back-dating and forward-dating may be important in some applications when anticipation or delayed effects are a concern.

The treatment indicator for each unit and time period is
\begin{align*}
    D_{i,t}(A_i) = \I(t  \geq A_i),
\end{align*}
which respects the temporal sequence of the real assignment: given an adoption time $a$, a treated period always occurs for $t\geq a$, while an untreated period always occurs for $t < a$. This mechanism imposes different restrictions on the matrix $\mathbf{D}$. For each unit $i$, there are now $J$ possible adoption times $a\in\mathcal{A}$ where $D_{i,t} = 0$ for $t<a$ and $D_{i,t} = 1$ for $t\geq a$. Since each row of $\mathbf{D}$ corresponds to the assignment of each unit, and each unit is assigned an adoption time independently, there are $J^{n}$ possible values of $\mathbf{D}$. Denoting by $\mathcal{D}_{\texttt{AT}}$ the set of all allowed assignment matrices, we have $\P[ \mathbf{D}  = \mathbf{d} ] = J^{-n}$ for $\mathbf{d} \in \mathcal{D}_{\texttt{AT}}$.

The conditions to apply randomization inference under this mechanism are summarized in the following assumption. 
\begin{assumption}[AT Assignment Mechanism]\label{ass:AM2}
    There exists a window $\mathcal{W}_\tau$ of $\tau$ periods before and after the time of the intervention such that the following conditions hold for all $i=1,2,\cdots,n$.
    \begin{enumerate}[label=\emph{(\roman*)},leftmargin=*]
        \item $\bar{y}_{i,\tau,-}(0,a)$, $\bar{y}_{i,\tau,-}(1,a)$, $\bar{y}_{i,\tau,+}(0,a)$ and $\bar{y}_{i,\tau,+}(1,a)$ are non-stochastic, for all $a\in \mathcal{A}$.
            
        \item $\P[ \mathbf{D}  = \mathbf{d} ] = J^{-n}$ for all $\mathbf{d} \in \mathcal{D}_{\texttt{AT}}$.
    \end{enumerate}
\end{assumption}

Assumption \ref{ass:AM2}(i) is analogous to Assumption \ref{ass:AM1}(i), while  Assumption \ref{ass:AM2}(ii) is analogous to Assumption \ref{ass:AM1}(iii). Under this assignment mechanism, a no-trends assumption is not needed.

We consider the following null hypothesis,
\begin{align*}
    \mathsf{H}_{\mathtt{AT}}: \;
    & \sum_{t\in\mathcal{W}_\tau} \I(a \leq t < \a_0) [ y_{i,t}(0) - y_{i,t}(1) ] = 0,\\
    & \sum_{t\in\mathcal{W}_\tau} \I(\a_0 \leq t < a) [ y_{i,t}(0) - y_{i,t}(1) ] = 0,\\
    &\text{for all $i=1,2,\ldots, n$ and $a\in\mathcal{A}$}.
\end{align*}

In contrast to the null hypothesis under the TR assignment mechanism, $\mathsf{H}_{\mathtt{AT}}$ is stated for the sum of individual potential outcomes for periods in $\mathcal{A}$ rather than average outcomes. This is necessary to impute the average potential outcomes $\bar{y}_{i,\tau,+}(1,a)$ and $\bar{y}_{i,\tau,+}(0,a)$ for all possible values of $a \in \mathcal{A}$. To see this, note that the average observed treated outcome is $\bar{Y}_{i,\tau,1} = \frac{1}{N_{i,\tau,1}}\sum_{t \in \mathcal{W}_{\tau}} D_{it} Y_{it} = \frac{1}{\sum_{t \in \mathcal{W}_{\tau}} \I(t \geq a_0) }\sum_{t \in \mathcal{W}_{\tau}} \I(t \geq a_0) y_{i,t}(1) = \bar{y}_{i,\tau,+}(1,\a_0)$. Whenever the assignment mechanism moves the adoption time from $\a_0$ to another value $a>\a_0$, the observations for periods $\a_0 \leq t <  a$ must be assigned to the control group, even though they are treated according to the actual mechanism. The second condition in $\mathsf{H}_{\mathtt{AT}}$ allows for the imputation of the untreated average potential outcome $\bar{y}_{i,\tau,+}(0,a)$ in this case. A similar argument shows that the first condition allows the imputation of $\bar{y}_{i,\tau,+}(1,a)$ when $a < \a_0$. If the set $\mathcal{A}$ only includes back-dating (forward-dating), only the first (second) condition in $\mathsf{H}_{\mathtt{AT}}$ will be applicable.

If $J=1$, $\mathsf{H}_{\mathtt{AT}}$ is equivalent to $y_{i,{\a_0-1}}(0) - y_{i,{\a_0-1}}(1) = 0$ and $y_{i,{\a_0}}(0) - y_{i,{\a_0}}(1) = 0$ for all $i=1,2,\ldots, n$ and $a\in\mathcal{A}$, which is a sharp null hypothesis for individual potential outcomes. For $J>1$, $\mathsf{H}_{\mathtt{AT}}$ is weaker than assuming no treatment effects for every unit in every period. The number of forward-dating periods included in $\mathcal{A}$ can be as large as $\tau$, while the number of back-dating periods can be as large as $\tau-1$ to ensure that at least one period is untreated for every unit. 

Under Assumption \ref{ass:AM2} and $\mathsf{H}_{\mathtt{AT}}$, once again  for any test statistic $S(\mathbf{D},\bar{\mathbf{Y}}_{0},\bar{\mathbf{Y}}_{1})$ we have $S(\mathbf{D},\bar{\mathbf{Y}}_{0},\bar{\mathbf{Y}}_{1}) = S(\mathbf{D},\bar{\mathbf{y}}(\a_0))$, and therefore its distribution  is fully determined by the randomization distribution of $\mathbf{D}$. The randomization-based p-value can be obtained analogously to $p_\mathtt{TR}$ as
\begin{align*}
    p_\mathtt{AT}
    &= \P\big[|S(\mathbf{D},\bar{\mathbf{Y}}_0,\bar{\mathbf{Y}}_1)| \geq |s_\mathtt{obs}| \big]
    =  \frac{1}{J^n} \sum_{\mathbf{d}\in\mathcal{D}_{\mathtt{AT}}} \I(|S(\mathbf{d},\bar{\mathbf{y}}(\a_0))| \geq |s_\mathtt{obs}|),
\end{align*}

\subsection{Other Inference Procedures}

The Fisherian framework can also be used to derive randomization-based confidence intervals by inversion of hypothesis tests under a model of treatment effects. We follow \cite{Rosenbaum_2002_book} and assume a constant treatment effect model. In the TR assignment mechanism, this modifies our null hypothesis to $\tilde{\mathsf{H}}_{\mathtt{TR}}: y_{i,\tau,+}(1)=y_{i,\tau,+}(0) + \theta_0$ for all $i=1,2,\ldots, n$. Confidence intervals of level $1-\alpha$ are obtained by collecting all the values of $\theta_0$ that are not rejected by a randomization-based test of  $\tilde{\mathsf{H}}_{\mathtt{TR}}$ at level $\alpha$. The calculation of our Fisherian p-value is the same as before, except that we use the adjusted observed outcomes instead of the raw observed outcomes, where the adjustment is conducted using the constant treatment effect model, i.e. $Y_{i,\tau,k,\mathtt{adj}}=Y_{i,\tau,k} - \theta_0 D_{i,t}$ for $k=0,1$. Similar ideas can be used for inversion of $\mathsf{H}_{\mathtt{AT}}$.

Our approach can also be generalized to a joint test of multiple hypotheses, which allows us to test hypotheses in many windows simultaneously around adoption time, rather than performing one test per window. The null hypothesis can be made for $L$ different windows $\mathcal{W}_{\tau_l}$, $l = 1,2,\ldots, L$, under either the TR or AT mechanisms. Then, a test statistic can be calculated for each hypothesis in each window following the steps above. Letting $s_{l}$ be the test statistic associated with window $\tau_l$, we can combine the statistics for $L$ windows, $s_1, s_2, \ldots, s_L$ into a single joint statistic such as the mean, maximum, or Hotelling's $T^2$, and calculate a p-value by repeating the procedure for different realizations of the treatment vector. This procedure can be used to test the hypothesis that the effect is jointly zero in all windows $\mathcal{W}_1, \mathcal{W}_2, \ldots, \mathcal{W}_L$, as we illustrate in the following section.

\subsection{Window Selection}\label{subsec:methods_Wsel}

Our hypothesized randomization mechanisms assume that the treatment assignment could have been different in a small window around adoption time. Moreover, the TR mechanism assumes that, within this window, there is no difference in average potential outcomes between the pre-adoption and post-adoption periods. The implementation of our approach thus requires the careful choice of $\tau$, the length of the window around adoption time. 

Our approach allows us to reduce extrapolation to a minimum by considering the smallest window $\mathcal{W}_\tau$ around adoption time. By choosing $\tau=1$, researchers can compare the average outcome in the period immediately before adoption of the intervention to the average outcome in the exact period when the intervention goes into effect. Although this window minimizes extrapolation as much as it is possible given the data, it may be too small for causal effects to appear. 

For windows $\mathcal{W}_\tau$ larger than the minimum, two types of strategies can guide the choice of $\tau$. The first strategy is qualitative, and it involves considering the mechanisms by which the intervention is likely to affect the outcome, and the period of time that is plausibly required for those mechanisms to become active and cause visible changes. In our application, it is in principle possible for the change in incentives brought about by the new CCP to cause individuals to change their criminal behavior the day of adoption. Thus, a window of one day before to one day after the intervention is reasonable in our context to capture immediate effects. However, in other applications, this window length might not be practical. For example, for scientists who wish to consider the effect of emission policies on climate change, a window of one day before and after the policy will be uninformative, as we know that the mechanisms by which reducing emissions affect climate take years to mature.  

The second strategy is quantitative, which we develop following the method by  \cite{Cattaneo-Frandsen-Titiunik_2015_JCI} for window selection in local randomization RD designs. In the original method, pre-intervention covariates (for which the treatment effect is known to be zero) are used to choose the window, sequentially decreasing its size until covariates are balanced and the null hypothesis of no treatment effect fails to be rejected. 

In our context, all pre-intervention covariates are balanced exactly in any window around adoption time, because we observe the same exact units before and after the intervention. Thus, a window selector procedure cannot be based on pre-intervention covariates. Instead of covariates, we propose a method based on the time-varying outcome before the intervention occurs. Because all units are untreated before the intervention, the null hypothesis of no treatment effect should fail to be rejected unless there are time trends that cause an incorrect rejection. The suggested procedure is therefore as follows: (i) pick a date before the actual intervention to be used as an artificial adoption time; (ii) choose a sequence of window lengths, for example from 1 until the end of the pre-treatment period, and obtain the randomization-based p-value for each window in this sequence; and (iii) select the window length such that the p-value is (approximately) larger than 0.15 in that window and in all windows of smaller length.

The procedure will work under the assumption that the time trends that characterize the outcome in the artificial windows are similar to the time trends that characterize the window around the real intervention. In the absence of this assumption, any balance in the outcome in artificial windows will be uninformative regarding the plausibility of the no trends assumption. This is similar to the role of covariate balance tests in randomized controlled experiments, where observing balance in covariates is taken as indirect evidence towards the plausibility of the independence between the treatment assignment and the potential outcomes. We illustrate how to implement this procedure in Section \ref{sec:results}. 

As in the RD case, the selection of the window based on the pre-intervention outcome is in itself a type of falsification analysis: if the window length is chosen following the outlined procedure and researchers assume that time trends are similar within the artificial and the real windows, the window selector suggests values of $\tau$ where the assumption of no trends might be plausible. In addition to the window selection procedure, our framework allows us to design other falsification tests based on prior periods. In our application, the intervention occurred on November 1st, 2017, but we have data on our outcome variable from several years prior. For our chosen window lengths using the window selection procedure, we consider additional falsification analyses that study the effect of the intervention in the years before and after the true intervention date, setting the cutoff to  artificial values. We discuss this procedure in Section \ref{sec:results}.

\subsection{Choice of Test Statistic}

The Fisherian framework provides exact inferences for any suitable test statistic whose distribution is known under the null. We use the average difference between the (average) outcome in the treated post-adoption period and the (average) outcome in the untreated pre-adoption period, $S(\mathbf{D},\bar{\mathbf{Y}}_0,\bar{\mathbf{Y}}_1) = \frac{1}{n} \sum_{i=1}^{n} \bar{Y}_{i,\tau,1} - \frac{1}{n} \sum_{i=1}^{n} \bar{Y}_{i,\tau,0} \equiv \widehat{\theta}_{\tau}$, which is justified by our assumptions and null hypotheses. 

However, the framework is general and allows for the use of other test statistics. In particular, a test statistic that adjusts for time trends could be useful in our context, as it could allow researchers to analyze larger windows by using the adjusted outcome instead of the raw outcome---see, e.g., \cite{Cattaneo-Titiunik-VazquezBare_2017_JPAM}. For this, researchers would need to modify Assumptions \ref{ass:AM1} and \ref{ass:AM2} to refer to the adjusted outcomes. In Section SA-1 in the Supplemental Appendix, we present results using a time-adjusted statistic.

\section{Results} \label{sec:results}

The new adversarial system came into effect nationwide at 12 a.m. on November 1st, 2017. The cornerstone of our research design is hypothesizing different assignment mechanisms for the reform affecting each of Montevideo's 62 neighborhoods in a small window around adoption time. We include in our analysis the smallest window, $\mathcal{W}_1$, which includes just two days and compares the offenses that occurred on October 31st, 2017, to the offenses that occurred on November 1st, 2017. The smallest possible window allows researchers to investigate whether treatment effects appear almost instantaneously. If the intervention requires more than one unit of time to affect the outcome, the effects in the smallest window will be null.

To guide the choice of additional windows, we implement the window selection procedure outlined in Section \ref{sec:methods}. We choose an artificial or `placebo' adoption time before the actual intervention, and use the TR mechanism to test the null hypothesis $\mathsf{H}_{\mathtt{TR}}$ that the placebo intervention had no effect on total crime for many windows around this artificial adoption time. Our placebo adoption time is the first Wednesday in October, 2017, about one month prior to the adoption of the real intervention, which occurred on the first Wednesday in November.

\begin{figure}[h]
\vspace{-0.4in}
\centering
\includegraphics[scale=0.55]{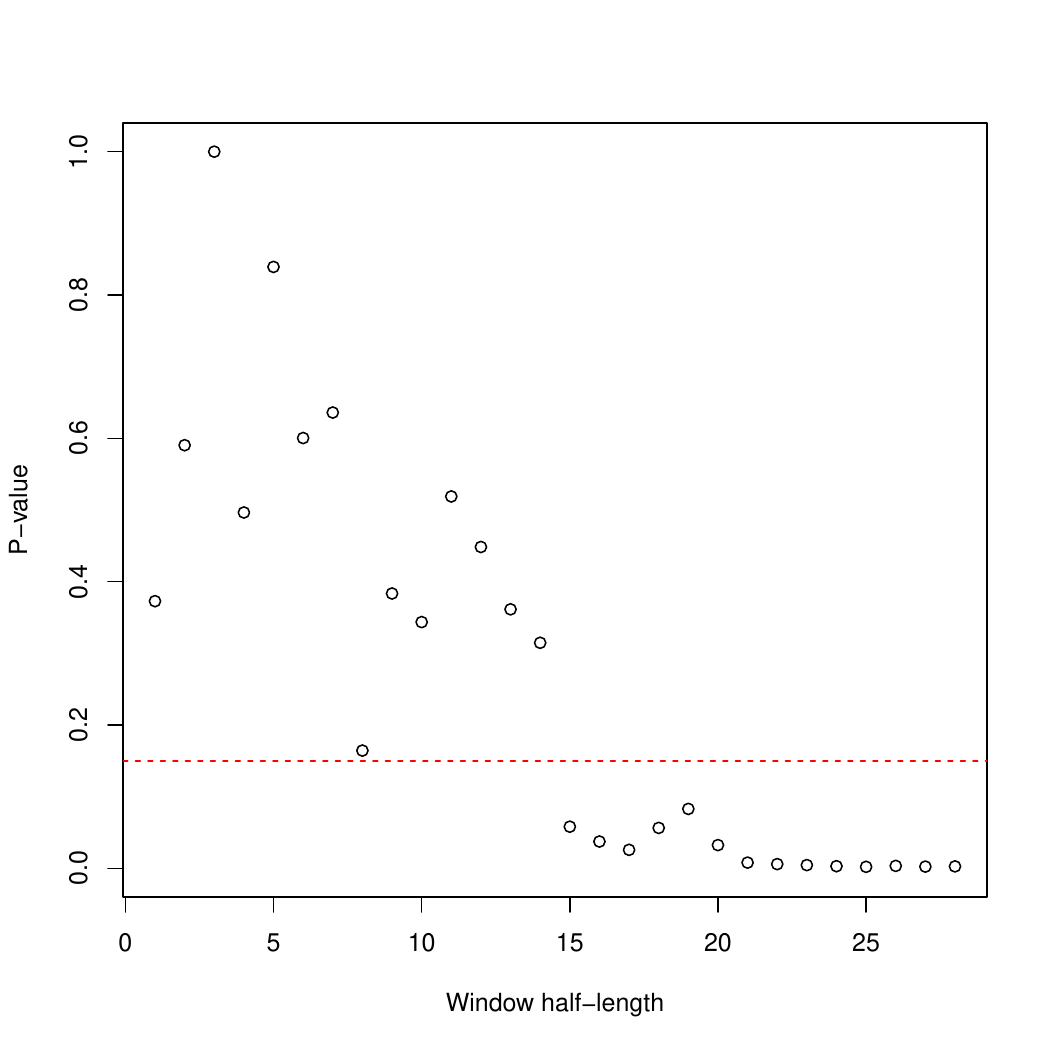} 
\caption{Window Selector Around {\it Placebo} Adoption Times\\
\footnotesize{{\it  Placebo adoption times}: Up to 28 days prior to November 1, 2017}\\
\footnotesize{{\it Outcome}: Daily number of crimes reported to police}
}
\label{fig:windowsel_C28}
\end{figure}

Figure \ref{fig:windowsel_C28} plots the randomization-based p-value against the half length $\tau$ of each window. We add a horizontal line at 0.15, following the recommendation of \citet{Cattaneo-Frandsen-Titiunik_2015_JCI}; since we are concerned with incorrectly failing to reject the null hypothesis (rather than incorrectly rejecting it), it is advisable to consider windows where failure to reject occurs for significance levels higher than conventional ones. The pattern shows that in all windows within 14 days of the selected placebo cutoff, the p-value associated with the null hypothesis $\mathsf{H}_{\mathtt{TR}}$ for total crime is 0.15 or higher, validating choices of $\tau \leq  14$. For larger windows, the p-value decreases rapidly---it is higher than 0.30 for $\tau=14$, less than 0.06 for $\tau=15$, and less than 0.01 for $\tau=21$. In other words, after day 15, the trend in the outcome is such that our randomization-based approach detects an effect even when there is none. This reinforces that our procedure is best suited for evaluating short-term effects.  

In Section SA-2 of the Supplemental Appendix, we repeat the exercise for two additional placebo adoption times (21 and 35 days before the actual adoption, $a=-21,-35$); the overall pattern is roughly similar, although some p-values fall below 0.15 for smaller $\tau$, particularly for $a=-21$.

Table \ref{tab:total} reports the results for our main outcome of interest, total crime reports, now using the real adoption time. The analysis is performed in three windows in the set of windows validated above: 1 day, 7 days, and 14 days before/after the adoption of the new code, that is, $\mathcal{W_\tau}$ for $\tau=1,7,14$. For the implementation of the AT mechanism, we consider backdating up to $\tau-1$ and include the actual adoption time: we set $\mathcal{A} = \{ \a_0-6,\a_0-5,\ldots, \a_0-1, \a_0\}$ for $\tau=7$, and $ \mathcal{A} = \{ \a_0-13,\a_0-12,\ldots, \a_0-1, \a_0\}$ for $\tau=14$. The table reports the estimated average difference, $\widehat{\theta}_{\tau}$, and the randomization-based p-value corresponding to the two-sided test of the null hypotheses $\mathsf{H}_{\mathtt{TR}}$ and $\mathsf{H}_{\mathtt{AT}}$, calculated using the TR and AT assignment mechanisms, respectively.

\vspace{10pt}
\begin{table}[htbp]
\caption{Short-Term Effects of CCP Reform in Montevideo \\
\small{{\it Outcome}: Daily number of crimes reported to police} \\
\small{{\it  Actual adoption time}: November 1, 2017}}

\label{tab:total}\centering
\par
{ \footnotesize
\begin{tabular}{ccccccc}
\toprule
\multirow{1}{*}{} &\multicolumn{2}{c}{Estimates} &\multicolumn{2}{c}{TR Mechanism}&\multicolumn{2}{c}{AT Mechanism} \\
\cmidrule(lr){2-3}\cmidrule(lr){4-5}\cmidrule(lr){6-7}
$\tau$&\multicolumn{1}{c}{$\widehat{\theta}_{\tau}$}  &\multicolumn{1}{c}{$\frac{1}{n} \sum_{i=1}^{n} \bar{Y}_{i,\tau,0}$} & p-value & 95\% CI & p-value & 95\% CI \\
\cmidrule(lr){1-1}\cmidrule(lr){2-2}\cmidrule(lr){3-3}\cmidrule(lr){4-5}\cmidrule(lr){6-7}
1 day & 0.839 & 5.306 & 0.131 & [-0.23, 1.90] & - & - \\ 
 7 days & 0.403 & 4.892 & 0.013 & [ 0.09, 0.72] & 0.430 & [-0.69, 0.87] \\ 
14 days & 0.298 & 5.065 & 0.020 & [ 0.05, 0.55] & 0.021 & [ 0.04, 1.81] \\ 
\bottomrule
 \multicolumn{7}{p{0.75\textwidth}}{ \scriptsize{Sample is 62 neighborhoods in Montevideo, each observed before and after the adoption of the CCP reform. The p-values are randomization-based for $\widehat{\theta}_{\tau} = \frac{1}{n} \sum_{i=1}^{n} \bar{Y}_{i,\tau,1} - \frac{1}{n} \sum_{i=1}^{n} \bar{Y}_{i,\tau,0}$ with 10,000 simulations. $\tau$ denotes the half-length of a symmetric window around adoption time. CI refers to confidence interval, calculated by inversion based on a constant treatment effect model. TR and AT refer to Treatment Reversal and Adoption Timing mechanisms, respectively. In the AT mechanism, $\mathcal{A} =\{-6,-5,\ldots,0\}$ for $\tau=7$ and $\mathcal{A} =\{-13,-12,\ldots,0\}$ for $\tau=14$}.}
\end{tabular}
}

\end{table}

The first row reports the results in the smallest possible window, $\mathcal{W}_1$. On the day before adoption, the average number of crime reports per neighborhood was 5.306; this average increased by 0.839 to 6.145 the day after the intervention, but the test of $\mathsf{H}_{\mathtt{TR}}$ based on the TR assignment mechanism does not reject the null at conventional levels (p-value is 0.131, 95\% confidence interval ranges from -0.23 through 1.9). Note that the AT mechanism cannot be used in $\mathcal{W}_1$ because it is not possible to vary the adoption time and still keep observations on both sides.

The second row shows that, when comparing the seven days before to the seven days after the intervention, the reform seems to have resulted in a statistically significant increase in the total number of crimes reported when using the TR mechanism (null hypothesis rejected with p-value 0.013, 95\% confidence interval of [0.09, 0.72] under a constant treatment effect model). The average number of daily crimes reported per neighborhood during the seven days preceding the intervention was 4.892, increasing to 5.295 in the week following the intervention---an increase of 0.403 police reports per day per neighborhood. Since there are 62 neighborhoods, this corresponds to an increase from approximately $4.892\times 62\times 7\approx 2{,}123$  reports in the seven days before the intervention to $(4.892+0.403)\times 62\times 7\approx 2{,}298$ after---an increase of about 8.2 percent. The pattern is similar for a window within 14 days of the start of the intervention---TR null $\mathsf{H}_{\mathtt{TR}}$ rejected with p-value 0.020, confidence interval [0.05, 0.55]. Using the AT mechanism, $\mathsf{H}_{\mathtt{AT}}$ is not rejected in $\mathcal{W}_7$ but is rejected in $\mathcal{W}_{14}$.

We also investigate the treatment effects in many windows simultaneously. Table \ref{tab:total-simul} shows the results of joint randomization-based tests for all windows between day $1$ and day $7$, $\mathcal{W}_1,\ldots, \mathcal{W}_7$, and all windows between day $1$ and day $14$, $\mathcal{W}_1,\ldots, \mathcal{W}_{14}$. The p-values are calculated using three test statistics: the maximum of the average difference in each window, Hotelling's $T^2$ statistic using the vector of average differences in each window with their respective covariance matrices, and the mean of the average difference in each window. When jointly considering the treatment effect in the first seven or the first fourteen windows under the TR mechanism, we reject $\mathsf{H}_{\mathtt{TR}}$ at 10\% or lower level in all cases but one. For example, according to the maximum average difference across all windows, the hypothesis of no effect has p-value $0.064$ for the first seven windows, and according to Hotelling's $T^2$, this p-value is 0.074. The p-values in the first 14 windows are smaller. The pattern is similar with the AT mechanism. Together, these joint hypothesis tests suggest a significant effect within 7 to 14 days after adoption of the new CCP.

\vspace{10pt}
\begin{table}[htbp]
\caption{Joint Inference for Short-Term Effects of CCP Reform in Montevideo \\
\small{{\it Outcome}: Daily number of crimes reported to police} \\
\small{{\it  Actual adoption time}: November 1, 2017}}
\label{tab:total-simul}\centering
\par
{ \footnotesize
\begin{tabular}{lrrrrrr}
\toprule
\multicolumn{7}{c}{TR Assignment Mechanism}\\
\multicolumn{7}{c}{Joint test of hypotheses $\mathsf{H}_{\mathtt{TR}}$, for $\tau=1,\ldots, K$}\\
\midrule
& \multicolumn{2}{c}{Max} & \multicolumn{2}{c}{Hotelling} & \multicolumn{2}{c}{Mean}\\
\cmidrule(lr){2-3}\cmidrule(lr){4-5}\cmidrule(lr){6-7}
&\multicolumn{1}{c}{Statistic}&\multicolumn{1}{c}{p-value}&\multicolumn{1}{c}{Statistic}&\multicolumn{1}{c}{p-value}&\multicolumn{1}{c}{Statistic}&\multicolumn{1}{c}{p-value}\\
\midrule
$\tau=1,\ldots, 7$ ($K=7$) & 0.839 & 0.064 & 12.995 & 0.074 & 0.292 & 0.172 \\ 
$\tau=1,\ldots, 14$ ($K=14$) & 0.839 & 0.064 & 37.748 & 0.004 & 0.302 & 0.059 \\ 
\midrule
\multicolumn{7}{c}{AT Assignment Mechanism}\\
\multicolumn{7}{c}{Joint test of hypotheses $\mathsf{H}_{\mathtt{AT}}$, for $\tau=2,\ldots, K$}\\
\midrule
& \multicolumn{2}{c}{Max} & \multicolumn{2}{c}{Hotelling} & \multicolumn{2}{c}{Mean}\\
\cmidrule(lr){2-3}\cmidrule(lr){4-5}\cmidrule(lr){6-7}
&\multicolumn{1}{c}{Statistic}&\multicolumn{1}{c}{p-value}&\multicolumn{1}{c}{Statistic}&\multicolumn{1}{c}{p-value}&\multicolumn{1}{c}{Statistic}&\multicolumn{1}{c}{p-value}\\
\midrule
$\tau=2,\ldots, 7$ ($K=7$) & 0.403 & 0.124 & 6.893 & 0.105 & 0.283 & 0.233 \\ 
$\tau=2,\ldots, 14$ ($K=14$) & 0.403 & 0.015 & 26.375 & 0.003 & 0.327 & 0.021 \\ 
\bottomrule
 \multicolumn{7}{p{0.8\textwidth}}{ \scriptsize{Sample is 62 neighborhoods in Montevideo, each observed before and after the adoption of the CCP reform. The p-values are randomization-based for $\widehat{\theta}_{\tau} = \frac{1}{n} \sum_{i=1}^{n} \bar{Y}_{i,\tau,1} - \frac{1}{n} \sum_{i=1}^{n} \bar{Y}_{i,\tau,0}$ with 10,000 simulations. $\tau$ denotes the half-length of a symmetric window around adoption time. TR and AT refer to Treatment Reversal and Adoption Timing mechanisms, respectively. In the AT mechanism, $\mathcal{A}=\{-K+1,-K+2,\ldots,0\}$.}}
\end{tabular}
}
\end{table}

\subsection{Assessing Assumptions}

We present additional results from a falsification analysis that analyzes our main outcome at artificial adoption times. Our approach is similar to the strategy of estimating effects at placebo cutoffs in the RD design \citep[see, e.g.][Section 5]{Cattaneo-Idrobo-Titiunik_2020_Book}, where the presence of significant effects at artificial cutoffs is interpreted as potentially casting doubt on the main identifying assumptions. 

We intend these analyses as an illustration of how to provide empirical support for the main assumptions, not as a formal test---the assumptions are fundamentally untestable, and the lack of effects at artificial adoption times is neither necessary nor sufficient for the assumptions to hold. Nonetheless, observing no effects at artificial adoption times can sometimes rule out some violations. For example, if another factor that affects crime other than the new CCP changes discontinuously between $t=\a_0-1$ and $t=\a_0$, the causal interpretation of the results would be incorrect. In our study, this could happen, for example, if $t=\a_0-1$ fell on a Sunday and $t=\a_0$ on a Monday, and crime were higher during the weekend (see \citeauthor{prieto2021}, \citeyear{prieto2021}, for evidence and discussion regarding the temporal concentration of crime). Observing no significant day-of-the-week effects in the absence of the intervention might dissipate fears about this potential violation.

Our falsification analyses are an example of negative control analyses or a placebo population test, where an experiment is repeated under conditions in which null results are expected and the researchers verify that a null result is indeed observed \citep{LipsitchTchetgenTchetgenCoen2010-Epid}. In general, for this type of analysis to be informative, researchers make the assumption that failure to obtain a null result implies a violation of the main assumptions invoked.

Table \ref{tab:total-placdate} presents results with the adoption time artificially set to midnight on November 1st for the years 2015, 2016, and 2018. The outcome, windows, and assignment mechanisms are identical to those used in Table \ref{tab:total}; the only change is the use of artificial adoption times. The results show that there are no distinguishable differences in crime reports around artificial adoption times in the same windows considered in Table \ref{tab:total}: the null hypothesis fails to be rejected in all cases except for $\tau=1$ in 2016 under the TR mechanism, but this corresponds to a negative value of test statistic (-0.855). 

In Section SA-3 in the Supplemental Appendix, we show that the falsification results in Table \ref{tab:total-placdate} remain robust when we consider alternative values of $\tau$, and also when we set the artificial adoption time to the same day of the week as the actual adoption (first Wednesday in November) rather than to the same date. We also report results for additional years.

\vspace{10pt}
\begin{table}[htbp]
\caption{Short-Term Effects of CCP Reform in Montevideo\\ Around {\it Placebo} Adoption Times (Date)\\
\small{{\it Outcome}: daily number of crimes reported to police} \\
\small{{\it  Actual adoption time}: November 1st, 2017}}
\label{tab:total-placdate}\centering
\par
{ \footnotesize
\begin{tabular}{crrrrrrrrr}
\toprule
 &\multicolumn{3}{c}{2015} &\multicolumn{3}{c}{2016}&\multicolumn{3}{c}{2018}\\
\cmidrule(lr){2-4}\cmidrule(lr){5-7}\cmidrule(lr){8-10}
  &&\multicolumn{2}{c}{p-value} &&\multicolumn{2}{c}{p-value}&&\multicolumn{2}{c}{p-value}\\
\cmidrule(lr){3-4}\cmidrule(lr){6-7}\cmidrule(lr){9-10}
$\tau$&\multicolumn{1}{c}{$\widehat{\theta}_{\tau}$ } &\multicolumn{1}{c}{TR}&\multicolumn{1}{c}{AT} &\multicolumn{1}{c}{$\widehat{\theta}_{\tau}$ } &\multicolumn{1}{c}{TR}&\multicolumn{1}{c}{AT} 
     &\multicolumn{1}{c}{$\widehat{\theta}_{\tau}$ } &\multicolumn{1}{c}{TR}&\multicolumn{1}{c}{AT}\\
\cmidrule(lr){1-1}\cmidrule(lr){2-4}\cmidrule(lr){5-7}\cmidrule(lr){8-10}
1 day & 0.048 & 0.934 & \multicolumn{1}{c}{-} & -0.871 & 0.036 & \multicolumn{1}{c}{-} & 0.661 & 0.170 & \multicolumn{1}{c}{-} \\ 
7 days & 0.090 & 0.544 & 0.879 & -0.041 & 0.806 & 0.761 & 0.002 & 1.000 & 0.994 \\ 
14 days & 0.015 & 0.914 & 0.878 & 0.035 & 0.761 & 0.921 & -0.113 & 0.401 & 0.468 \\ 
\bottomrule
 \multicolumn{10}{p{0.75\textwidth}}{ \scriptsize{Sample is 62 neighborhoods in Montevideo, each observed before and after the adoption of the CCP reform. The p-values are randomization-based for $\widehat{\theta}_{\tau} = \frac{1}{n} \sum_{i=1}^{n} \bar{Y}_{i,\tau,1} - \frac{1}{n} \sum_{i=1}^{n} \bar{Y}_{i,\tau,0}$ with 10,000 simulations. $\tau$ denotes the half-length of a symmetric window around adoption time. TR and AT refer to Treatment Reversal and Adoption Timing mechanisms, respectively. In AT mechanism, $\mathcal{A} =\{-6,-5,\ldots,0\}$ for $\tau=7$ and $\mathcal{A} =\{-13,-12,\ldots,0\}$ for $\tau=14$}.}
\end{tabular}
}
\end{table}

In sum, the rejections that we see in small windows around the actual adoption time are not seen when we consider windows of the same length around the same date or time in prior years. Moreover, because our analysis is based on the exact same units before and after adoption in all falsification and actual analyses, the results in Table \ref{tab:total} cannot be explained by any time-constant unit-specific characteristics or `fixed effects'.

\subsection{Potential Mechanisms} \label{policy}

As shown in Table \ref{tab:total}, the new CCP led to an average rise of 25 police reports per day ($\approx 0.403 \times 62$) in Montevideo during the week following adoption. The CCP reform had multiple components. The new system made plea bargaining easier, reduced pretrial detention, and affected coordination between the police and prosecutors, all of which could have plausibly affected crime rates through decreases in deterrence and/or incapacitation. Moreover, the new system could have affected whether and how individuals report crime, how the police patrol neighborhoods, and how crime data is collected. 

We start by considering the possibility that the effects we see reflect a change in how crime was reported rather than an actual deterioration in public safety. Although we cannot rule this out entirely, we have found no compelling evidence of a change in crime reporting practices. First, the reform did not alter how crime data were collected, managed, or disseminated; all processes continued to be overseen by the Ministry of the Interior through its Public Security Management System (SGSP). As part of the implementation of the new CCP, SGSP data began to be shared in real time with the Office of the Attorney General through the Uruguayan Adversarial Criminal Procedure Information System (SIPPAU). The connection between SGSP and SIPPAU reflected prosecutors' expanded role in the investigative and judicial stages of criminal proceedings under the new adversarial system. We have seen no evidence that this data transfer undermined the credibility of the crime statistics produced by the Ministry of the Interior.

Moreover, institutional trust trends do not support a sharp increase in the willingness to report crimes. According to \cite{latinobarometro2016, latinobarometro2018}, trust in the police remained stable at elevated levels: around 60\% of respondents reported ``some'' or ``a lot'' of trust in both 2016 and 2018. This is the highest level observed in Latin America, where the 2018 regional average stood at just 35\%. In contrast, trust in the judiciary declined by 8 percentage points, from 47\% in 2016 to 39\% in 2018, possibly reflecting dissatisfaction with procedural changes, such as limits on pretrial detention and the rollout of alternative procedures to oral trial. If anything, this decline could have discouraged reporting, suggesting our estimates may understate the reform's true impact. Moreover, in Section SA-4 of the Supplemental Appendix, we show that the increase in crime reports does not appear to have been driven by enhanced street-level policing. 

We believe our estimates reflect a rise in actual crime driven the various components of the bundle of interventions introduced by the new CCP. The reform may have led to less severe sanctions and a lower probability of conviction, factors that have been shown to increase crime. First, the new system seems to have resulted in less severe penalties, as it introduced procedural alternatives to oral trials that are associated with lighter sentences. These alternatives, which were employed in 90\% of solved cases, help lawyers and defendants resolve their cases faster while allowing public prosecutors to avoid lengthy and demanding criminal trials. Most of these solved cases are the result of an abbreviated process (i.e., a type of plea bargaining introduced by the new CCP) that implies an agreement between the defendant and the prosecutor whereby the former pleads guilty to a particular charge in return for a more lenient sentence. Moreover, preventive detention (i.e., detention while the process lasts until there is a sentence) ceased to be the norm, in contrast to the old system, which might have created the expectation of less severe punishment. Second, the reform might have resulted in a lower probability of conviction. Under the new system, prosecutors faced a significant increase in their workload, as they were now exclusively responsible for leading the investigation and carrying the evidence to judges. Moreover, the police must conduct investigations under new supervisors (prosecutors instead of judges) and different rules, which could have created coordination challenges.

\begin{figure}[!ht]
\subfloat[Police Reports and Prison Population]{\includegraphics[width =4.5in]{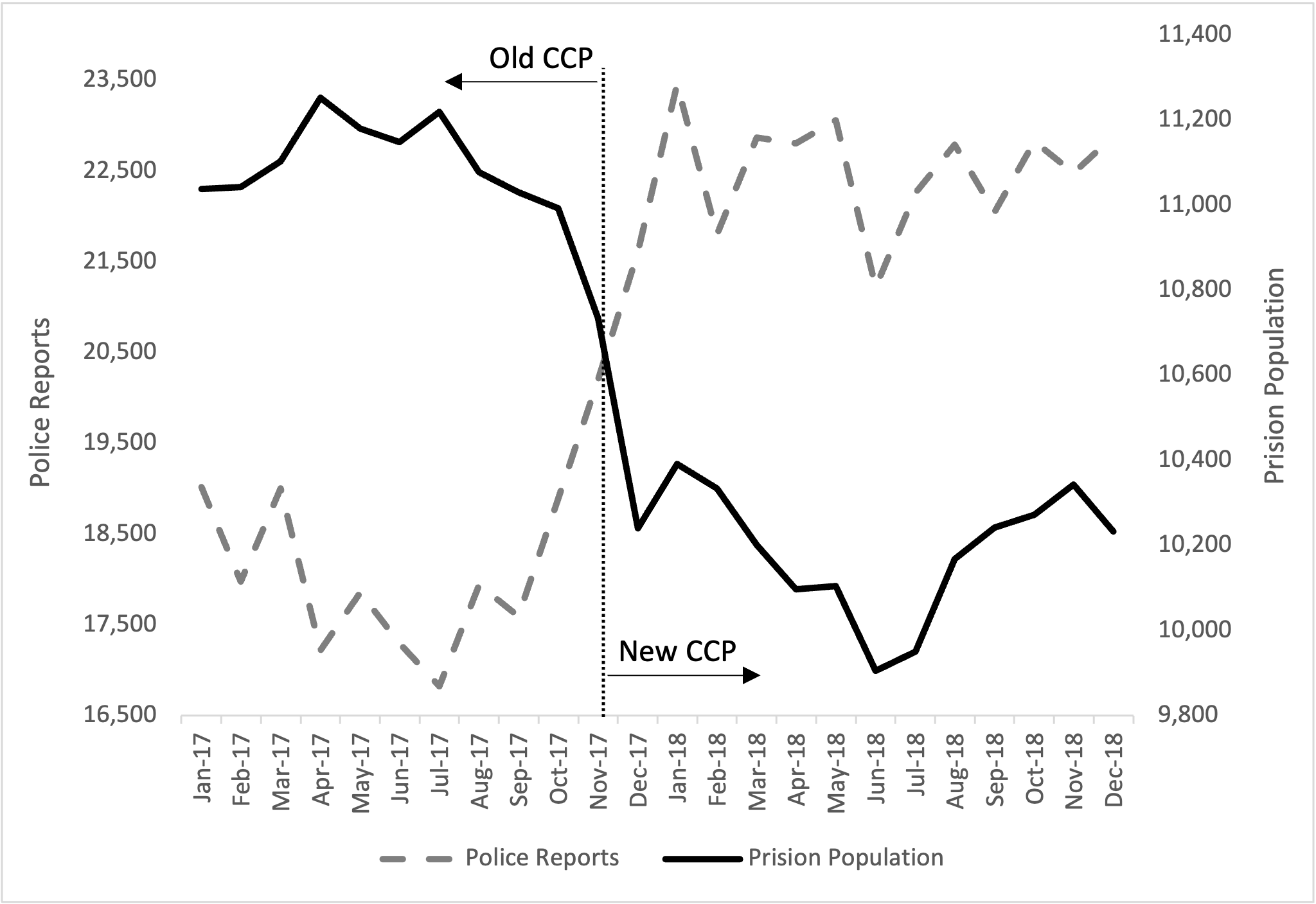}} \\
\subfloat[Criminal Imputations]{\includegraphics[width = 4.5in]{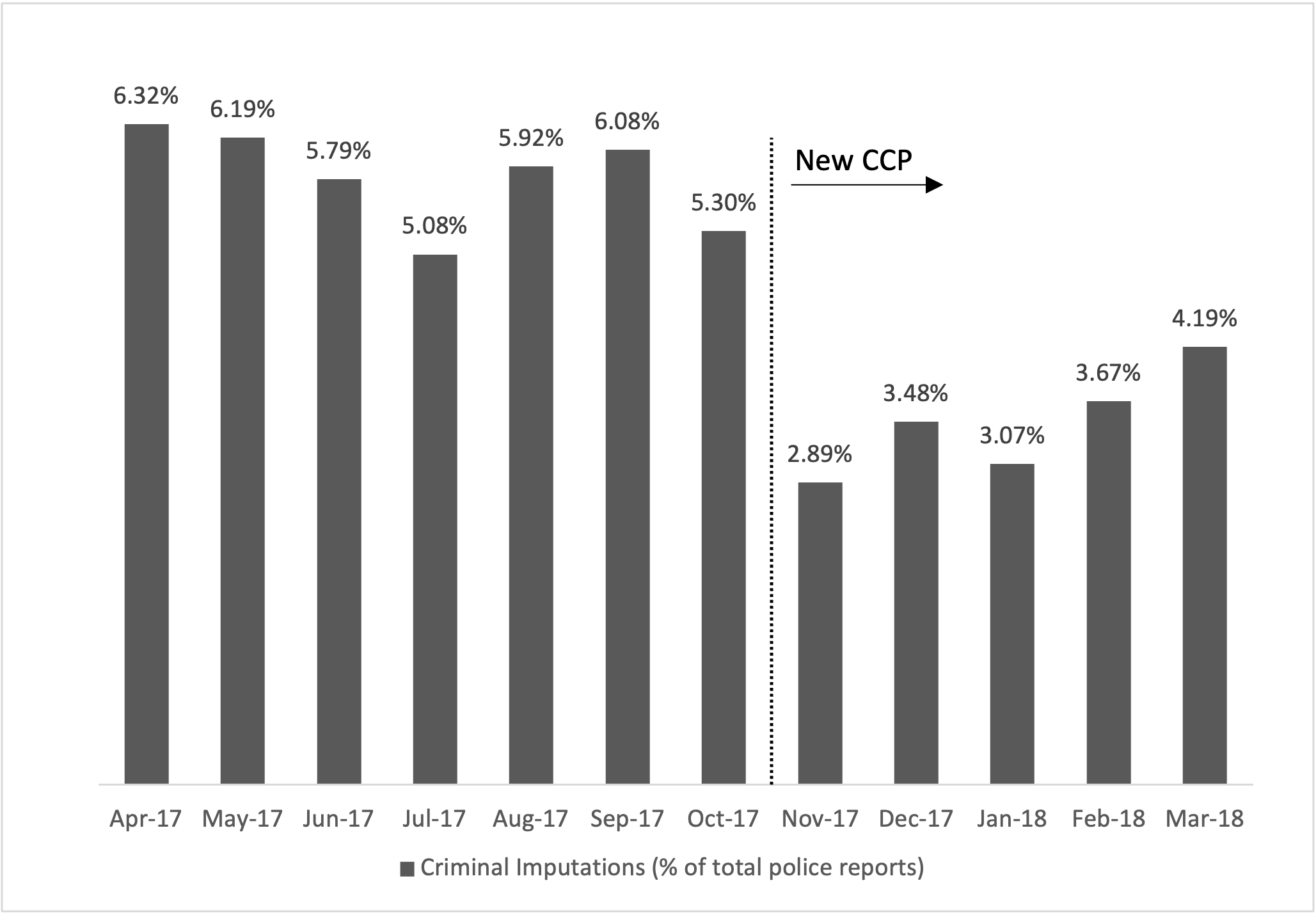}}
\caption{Police Reports, Prison Population and Criminal Imputations}
\label{F.05}
\end{figure}

Figure~\ref{F.05} provides some evidence supporting the hypothesis that the reform affected crime incentives. For the first time in more than a decade, the average number of people in prisons decreased in 2018, from 11,005 to 10,179 inmates (a 7.5\% decrease). According to Figure~\ref{F.05}(a), this sharp reduction in prison population starts at the same time as the implementation of the new CCP and contrasts with the sudden rise in the number of crimes reported to police that we documented in the prior section. Meanwhile, the number of criminal indictments (i.e., formal accusations made by public prosecutors) also experienced a strong month-to-month decrease in November 2017, from 1,001 to 584 cases (i.e., a 42\% reduction). Figure~\ref{F.05}(b) illustrates the evolution of indictments relative to police reports. The average ratio for the first two months of the new CCP is 3.2\% (i.e., November and December 2017), well below its value of 5.9\% during the rest of 2017 in the last months of the old CCP. 

The above trends suggest that less severe crimes could have increased due to a change in crime incentives (i.e., less deterrence) and also to a rise in the number of active criminals (i.e., less incapacitation). The reduction in prison population might be explained by the new constraints imposed on the use of preventive detention by prosecutors. As mentioned above, preventive prison was used extensively under the old system, in particular for cases of recidivism (55\% of the cases). This was the norm for both property crimes (e.g., thefts) and violent crimes (e.g., robberies and domestic violence). However, under the new system, preventive detention is applicable only when there is sufficient evidence that the defendant might attempt to escape, obstruct the investigation, or pose a risk to society (i.e., severe and typically violent crimes). Consequently, a convicted offender who re-offends by committing a lesser crime could be immediately released under the new CCP, whereas under the old CCP, they would likely have served preventive detention. 

If the increase in crime reports is due to a more selective use of preventive detention, we should not observe any impact on domestic violence (a crime considered very severe by prosecutors under both procedural regimes),  while the effects should be present for thefts, as they are non-violent. In Section SA-4 of the Supplemental Appendix, we report disaggregated effects by type of offense and show that the overall effects are not driven by the the two most frequent violent crimes (domestic violence and robberies).

\section{Conclusion}

We developed a randomization-based framework for short-term causal inference, designed to analyze before-and-after event studies where an intervention is given to all cross-sectional units at the same time, and all units are observed for several periods before and after the intervention. Our setup assumes that the units' potential outcomes are non-stochastic, and considers two possible assignment mechanisms governing how units receive the treatment in a small window around the time of adoption of the intervention. We used this framework to study the effects of a reform to the code of criminal procedure introduced in Uruguay on November 1, 2017, which was simultaneously implemented in the entire country. The reform increased total crime reports in Montevideo by about $8.2$ percent in the week immediately after adoption. Our randomization-based framework rejects hypotheses of no effect in small windows around adoption time, providing evidence that the reform increased crime reports. Moreover, we see no evidence of such increase when the analysis is conducted in prior years, before the intervention is active.

Our framework is tailored to study immediate or short-term effects, which are essential to the practical implementation of institutional reforms. Our case study illustrates how a shift in the criminal adjudication system can trigger immediate and unintended consequences that materially shape the reform's long-term trajectory. While the full impact of reforms can only be assessed over a medium- to long-term horizon, early signals play a critical role in shaping public opinion and securing political support. From a public policy perspective, our findings underscore the importance of embedding institutional change within a framework of real-time empirical monitoring. Our proposed framework was designed specifically for short-term causal inference, facilitating evidence-based monitoring of reform outcomes from the outset. By enabling policymakers to detect early shocks, diagnose implementation challenges, and adjust course in real time, this approach enhances the capacity to steer complex reforms toward long-term success.

\onehalfspacing
\bibliographystyle{jasa}
\bibliography{CDT_2025_JRSSA--bib}

\begin{thebibliography}{4}
\newcommand{\enquote}[1]{``#1''}
\expandafter\ifx\csname natexlab\endcsname\relax\def\natexlab#1{#1}\fi

\bibitem[Freyaldenhoven {\normalfont et~al.}(2025{\natexlab{a}})Freyaldenhoven,
  Hansen, P{\'e}rez and Shapiro]{FreyaldenhovenEtal2024-wp}
Freyaldenhoven, S., Hansen, C., P{\'e}rez, J.~P., and Shapiro, J.~M.
  (2025{\natexlab{a}}), \enquote{Visualization, identification, and estimation
  in the linear panel event-study design,} \emph{Advances in Economics and
  Econometrics: Twelfth World Congress}.
\bibitem[Freyaldenhoven {\normalfont et~al.}(2019)Freyaldenhoven, Hansen and
  Shapiro]{FreyaldenhovenEtal2019-AER}
Freyaldenhoven, S., Hansen, C., and Shapiro, J.~M. (2019), \enquote{Pre-event
  trends in the panel event-study design,} \emph{American Economic Review},
  109, 3307--3338.
\bibitem[Freyaldenhoven {\normalfont et~al.}(2025{\natexlab{b}})Freyaldenhoven,
  Hansen, P{\'e}rez, Shapiro and Carreto]{FreyaldenhovenEtal2025-Stata}
Freyaldenhoven, S., Hansen, C.~B., P{\'e}rez, J.~P., Shapiro, J.~M., and
  Carreto, C. (2025{\natexlab{b}}), \enquote{xtevent: Estimation and
  visualization in the linear panel event-study design,} \emph{The Stata
  Journal}, 25, 97--135.
\bibitem[Hausman and Kronick(2023)]{Hausman_Kronick}
Hausman, D., and Kronick, D. (2023), \enquote{The illusory end of stop and
  frisk in Chicago?} \emph{Science advances}, 9, eadh3017.
\end{thebibliography}


\begin{thebibliography}{36}
\newcommand{\enquote}[1]{``#1''}
\expandafter\ifx\csname natexlab\endcsname\relax\def\natexlab#1{#1}\fi

\bibitem[Acemoglu and Johnson(2005)]{acemoglu2005unbundling}
Acemoglu, D., and Johnson, S. (2005), \enquote{{Unbundling Institutions},}
  \emph{Journal of Political Economy}, 113, 949--995.
\bibitem[Atkins and Rubin(2003)]{atkins2003effects}
Atkins, R.~A., and Rubin, P.~H. (2003), \enquote{{Effects of Criminal Procedure
  on Crime Rates: Mapping Out the Consequences of the Exclusionary Rule},}
  \emph{The Journal of Law and Economics}, 46, 157--179.
\bibitem[Becker(1968)]{Becker_1968}
Becker, G.~S. (1968), \enquote{{Crime and Punishment: An Economic Approach},}
  \emph{Journal of Political Economy}, 76, 169--217.
\bibitem[Bushway and Redlich(2012)]{bushway2012plea}
Bushway, S.~D., and Redlich, A.~D. (2012), \enquote{{Is Plea Bargaining in the
  ``Shadow of the Trial'' a Mirage?}} \emph{Journal of Quantitative
  Criminology}, 28, 437--454.
\bibitem[Cattaneo {\normalfont et~al.}(2015)Cattaneo, Frandsen and
  Titiunik]{Cattaneo-Frandsen-Titiunik_2015_JCI}
Cattaneo, M.~D., Frandsen, B., and Titiunik, R. (2015), \enquote{{Randomization
  Inference in the Regression Discontinuity Design: An Application to Party
  Advantages in the U.S. Senate},} \emph{Journal of Causal Inference}, 3,
  1--24.
\bibitem[Cattaneo {\normalfont et~al.}(2020)Cattaneo, Idrobo and
  Titiunik]{Cattaneo-Idrobo-Titiunik_2020_Book}
Cattaneo, M.~D., Idrobo, N., and Titiunik, R. (2020), \emph{A Practical
  Introduction to Regression Discontinuity Designs: Foundations}, Cambridge
  Elements: Quantitative and Computational Methods for Social Science,
  Cambridge University Press.
\bibitem[Cattaneo {\normalfont et~al.}(2024)Cattaneo, Idrobo and
  Titiunik]{Cattaneo-Idrobo-Titiunik_2024_Book}
\leavevmode\vrule height .65ex depth -.6ex width 3em\  (2024), \emph{A
  Practical Introduction to Regression Discontinuity Designs: Extensions},
  Cambridge Elements: Quantitative and Computational Methods for Social
  Science, Cambridge University Press.
\bibitem[Cattaneo and Titiunik(2022)]{Cattaneo-Titiunik_2022_ARE}
Cattaneo, M.~D., and Titiunik, R. (2022), \enquote{Regression discontinuity
  designs,} \emph{Annual Review of Economics}, 14, 821--851.
\bibitem[Cattaneo {\normalfont et~al.}(2017)Cattaneo, Titiunik and
  Vazquez-Bare]{Cattaneo-Titiunik-VazquezBare_2017_JPAM}
Cattaneo, M.~D., Titiunik, R., and Vazquez-Bare, G. (2017), \enquote{{Comparing
  Inference Approaches for RD Designs: A Reexamination of the Effect of Head
  Start on Child Mortality},} \emph{Journal of Policy Analysis and Management},
  36, 643--681.
\bibitem[Cook {\normalfont et~al.}(2002)Cook, Campbell and
  Shadish]{ShadishCookCampbell2002-book}
Cook, T.~D., Campbell, D.~T., and Shadish, W. (2002), \emph{Experimental and
  quasi-experimental designs for generalized causal inference}, Houghton
  Mifflin.
\bibitem[Dalla~Pellegrina(2008)]{Dalla}
Dalla~Pellegrina, L. (2008), \enquote{{Court Delays and Crime Deterrence},}
  \emph{European Journal of Law and Economics}, 26, 267--290.
\bibitem[Du\^sek(2015)]{Dusek}
Du\^sek, L. (2015), \enquote{{Time to Punishment: The Effects of a Shorter
  Criminal Procedure on Crime Rates},} \emph{International Review of Law and
  Economics}, 43, 134--147.
\bibitem[Fandi{\~n}o and Gonz{\'a}lez~Postigo(2020)]{fandino2020adversarial}
Fandi{\~n}o, M., and Gonz{\'a}lez~Postigo, L. (2020), \enquote{{Adversarial
  Criminal Justice in Latin America: Comparative Analysis and Proposals},}
  \emph{Justice Studies Center of the Americas (JSCA)}.
\bibitem[Freyaldenhoven {\normalfont et~al.}(2025)Freyaldenhoven, Hansen,
  P{\'e}rez and Shapiro]{FreyaldenhovenEtal2024-wp}
Freyaldenhoven, S., Hansen, C., P{\'e}rez, J.~P., and Shapiro, J.~M. (2025),
  \enquote{Visualization, identification, and estimation in the linear panel
  event-study design,} \emph{Advances in Economics and Econometrics: Twelfth
  World Congress}.
\bibitem[Freyaldenhoven {\normalfont et~al.}(2019)Freyaldenhoven, Hansen and
  Shapiro]{FreyaldenhovenEtal2019-AER}
Freyaldenhoven, S., Hansen, C., and Shapiro, J.~M. (2019), \enquote{Pre-event
  trends in the panel event-study design,} \emph{American Economic Review},
  109, 3307--3338.
\bibitem[Hausman and Rapson(2018)]{HausmanRapson2018-AnnRevResEcon}
Hausman, C., and Rapson, D.~S. (2018), \enquote{Regression discontinuity in
  time: Considerations for empirical applications,} \emph{Annual Review of
  Resource Economics}, 10, 533--552.
\bibitem[Ho and Imai(2006)]{Ho-Imai_2006_JASA}
Ho, D.~E., and Imai, K. (2006), \enquote{Randomization Inference with Natural
  Experiments: An Analysis of Ballot Effects in the 2003 Election,}
  \emph{Journal of the American Statistical Association}, 101, 888--900.
\bibitem[Imbens and Rosenbaum(2005)]{ImbensRosenbaum_2005_jrssA}
Imbens, G.~W., and Rosenbaum, P. (2005), \enquote{Robust, accurate confidence
  intervals with a weak instrument: Quarter of birth and education,}
  \emph{Journal of the Royal Statistical Society, Series A}, 168, 109--126.
\bibitem[Jain and Mukand(2005)]{jain2005public}
Jain, S., and Mukand, S. (2005), \enquote{Public opinion and the dynamics of
  reform,} in \emph{Sixth Jacques Polak Annual Research Conference}.
\bibitem[Kang {\normalfont et~al.}(2018)Kang, Peck and
  Keele]{kang2018inference}
Kang, H., Peck, L., and Keele, L. (2018), \enquote{Inference for instrumental
  variables: a randomization inference approach,} \emph{Journal of the Royal
  Statistical Society Series A: Statistics in Society}, 181, 1231--1254.
\bibitem[Kronick(2019)]{Kronick}
Kronick, D. (2019), \enquote{{The Legal Origins of State Violence},}
  \emph{Unpublished Manuscript}.
\bibitem[Langer(2007)]{Langer}
Langer, M. (2007), \enquote{{Revolution in Latin American Criminal Procedure:
  Diffusion of Legal Ideas from the Periphery},} \emph{The American Journal of
  Comparative Law}, 55, 617--676.
\bibitem[{Latinobarómetro}(2016)]{latinobarometro2016}
{Latinobarómetro} (2016), \enquote{{Latinobarómetro Report 2016: Results by
  Sex and Age - Uruguay},}
  \url{https://www.latinobarometro.org/latContents.jsp}, Study LAT-2016.
  Accessed July 21, 2025.
\bibitem[{Latinobarómetro}(2018)]{latinobarometro2018}
\leavevmode\vrule height .65ex depth -.6ex width 3em\  (2018),
  \enquote{{Latinobarómetro Report 2018: Results by Sex and Age - Uruguay},}
  \url{https://www.latinobarometro.org}, Study LAT-2018. Accessed July 21,
  2025.
\bibitem[Lipsitch {\normalfont et~al.}(2010)Lipsitch, Tchetgen~Tchetgen and
  Cohen]{LipsitchTchetgenTchetgenCoen2010-Epid}
Lipsitch, M., Tchetgen~Tchetgen, E., and Cohen, T. (2010), \enquote{Negative
  controls: a tool for detecting confounding and bias in observational
  studies,} \emph{Epidemiology}, 21, 383--388.
\bibitem[Maier and Struensee(2000)]{maier2000introduccion}
Maier, J., and Struensee, E. (2000), \enquote{{Introducci{\'o}n: Las Reformas
  Procesales Penales en Am{\'e}rica Latina},} \emph{J. Maier, K. Ambos, \& J.
  Woischnik (Edits.), Las Reformas Procesales Penales en Am{\'e}rica Latina},
  17--34.
\bibitem[Miller(2023)]{Miller2023-JEP}
Miller, D.~L. (2023), \enquote{An introductory guide to event study models,}
  \emph{Journal of Economic Perspectives}, 37, 203--230.
\bibitem[Palumbo(2018)]{Palumbo}
Palumbo, I. (2018), \enquote{{Para Bonomi, el nuevo código trajo aparejado
  contradicciones entre jueces, fiscales y polic{\'\i}as},} \emph{Semanario
  Crónicas}, July 27th.
\bibitem[Prieto~Curiel(2021)]{prieto2021}
Prieto~Curiel, R. (2021), \enquote{{Weekly Crime Concentration},} \emph{Journal
  of Quantitative Criminology}, 1--28.
\bibitem[Rosenbaum(2002)]{Rosenbaum_2002_book}
Rosenbaum, P.~R. (2002), \emph{Observational Studies} (2nd ed.), New York:
  Springer.
\bibitem[Rosenbaum(2010)]{Rosenbaum2010-book}
Rosenbaum, P.~R. (2010), \emph{Design of Observational Studies}, Vol.~10,
  Springer.
\bibitem[Soares and Sviatschi(2010)]{Soares}
Soares, Y., and Sviatschi, M.~M. (2010), \enquote{{Does Court Efficiency Have a
  Deterrent Effect on Crime? Evidence for Costa Rica},} \emph{Unpublished
  Manuscript}.
\bibitem[Solomita(2019)]{solomita}
Solomita, M. (2019), \enquote{{Fiscales Agobiados: Turnos Interminables,
  Licencias por Estr{\'e}s y Jubilaciones Anticipadas},} \emph{Diario El
  Pa{\'\i}s}, Suplemento Qu{\'e} Pasa, May 18th.
\bibitem[Stokes(1996)]{stokes1996public}
Stokes, S.~C. (1996), \enquote{Public opinion and market reforms: the limits of
  economic voting,} \emph{Comparative Political Studies}, 29, 499--519.
\bibitem[Stokes(2001)]{stokes2001public}
Stokes, S.~C. (ed.) (2001), \emph{Public support for market reforms in new
  democracies}, Cambridge University Press.
\bibitem[Zorro~Medina {\normalfont et~al.}(2020)Zorro~Medina, Acosta and
  Mejia]{Zorro}
Zorro~Medina, A., Acosta, C., and Mejia, D. (2020), \enquote{{The Unintended
  Consequences of the U.S. Adversarial Model in Latin American Crime},}
  \emph{SSRN}.
\end{thebibliography}

\end{document}


\singlespacing

\title{
    \vspace{-0.5in} Randomization Inference for Before-and-After Studies with Multiple Units: An Application to a Criminal Procedure Reform in Uruguay \\\vspace{0.5in} Supplemental Appendix \bigskip 
}
\author{
    Matias D. Cattaneo\thanks{Department of Operations Research and Financial Engineering, Princeton University.}\and
    Carlos D\'iaz\thanks{Millennium Nucleus on Criminal Complexity, and
School of Economics and Business, Universidad Alberto Hurtado.}\and
    Roc\'io Titiunik\thanks{Department of Politics, Princeton University.}
}

\date{\today }

\maketitle
\thispagestyle{empty}
\clearpage
\doublespacing
\setcounter{page}{1}


\onehalfspacing
\setcounter{page}{1}
\pagestyle{plain}

\setcounter{tocdepth}{2}
\setcounter{secnumdepth}{4}
\tableofcontents

\clearpage

\section{Results with Time-Adjusted Test Statistic} 
\label{SAsec:timeadj}

In the main paper, we presented Fisherian randomization results using the unadjusted difference-in-means as the test statistic. We now present results using an alternative test statistic that removes the time trends in the  outcome using a linear specification. We fit a linear model of the outcome on a unit-specific intercept and a unit-specific linear time trend, using a window of 300 hundred days before and after the adoption of the real intervention for each unit. The time-adjusted test statistic is the difference in the average residuals of this fit between treatment and control groups. This effectively produces a de-trended version of the outcome. The interpretation of these results under our framework requires that we invoke Assumptions 1 and 2 for the adjusted outcomes. The results are presented in Table \ref{tab:totaladj}, employing the TR assignment mechanism.

\vspace{10pt}
\begin{table}[htbp]
\caption{Short-Term Effects of CCP Reform in Montevideo \\
\small{{\it Outcome}: Daily number of crimes reported to police} \\
\small{{\it  Actual adoption time}: November 1, 2017}}
\label{tab:totaladj}\centering
\par
{ \footnotesize
\begin{tabular}{ccccc}
\toprule
\multirow{1}{*}{} &\multicolumn{2}{c}{Estimates} &\multicolumn{2}{c}{TR Mechanism}\\
\cmidrule(lr){2-3}\cmidrule(lr){4-5} 
$\tau$&\multicolumn{1}{c}{$\widehat{\theta}_{\tau}$}  &\multicolumn{1}{c}{$\frac{1}{n} \sum_{i=1}^{n} \bar{Y}_{i,\tau,0}$} & p-value & 95\% CI\\
\cmidrule(lr){1-1}\cmidrule(lr){2-2}\cmidrule(lr){3-3}\cmidrule(lr){4-5}
1 day & 0.835 & 0.006 & 0.131 & [-0.22, 1.90] \\ 
  7 days & 0.375 & -0.397 & 0.021 & [ 0.06, 0.69] \\ 
  14 days & 0.243 & -0.210 & 0.055 & [ 0.00, 0.49] \\ 
\bottomrule\\
 \multicolumn{5}{p{0.5\textwidth}}{ \scriptsize{Sample is 62 neighborhoods in Montevideo, each observed before and after the adoption of the CCP reform. The p-values are randomization-based for $\widehat{\theta}_{\tau} = \frac{1}{n} \sum_{i=1}^{n} \bar{Y}_{i,\tau,1} - \frac{1}{n} \sum_{i=1}^{n} \bar{Y}_{i,\tau,0}$ with 10,000 simulations. $\tau$ denotes the half-length of a symmetric window around adoption time. CI refers to confidence interval, calculated by inversion based on a constant treatment effect model. TR refers to Treatment Reversal mechanism.}}
\end{tabular}
}
\end{table}

The advantage of using time-adjusted outcomes is that it potentially allows researchers to consider larger windows, as the assumption of no trends in the TR mechanism is more plausible for the de-trended outcome. The disadvantage is that it requires the assumption that the functional form used to model the outcome trend is a good enough approximation. In real applications, it may not be easy to assess how plausible a given adjustment model is. Nonetheless, the extension to time-adjusted test statistics makes our framework more general and will be relevant in some applications. 

\newpage
\section{Additional Window Selection Results}
\label{SAsec:windowsel}

We present the window selector described in the paper using the TR assignment mechanism, with two additional artificial adoption times: $a=-21$ (21 days before the actual adoption), and $a=-35$ (35 days before the actual adoption). The x-axis contains $\tau$, the half-length of the window $\mathcal{W}_\tau$, while the y-axis shows the p-value associated with a test of the Fisherian sharp null hypothesis that the treatment effect in the corresponding window $\mathcal{W}_\tau$ is zero for all units, using the TR assignment mechanism for the total crime outcome. The horizontal line is drawn at 0.15. 

\vspace{10pt}
\begin{figure}[h]
\vspace{-0.1in}
\centering
\begin{subfigure}{.48\textwidth}
\centering
\includegraphics[scale=0.45]{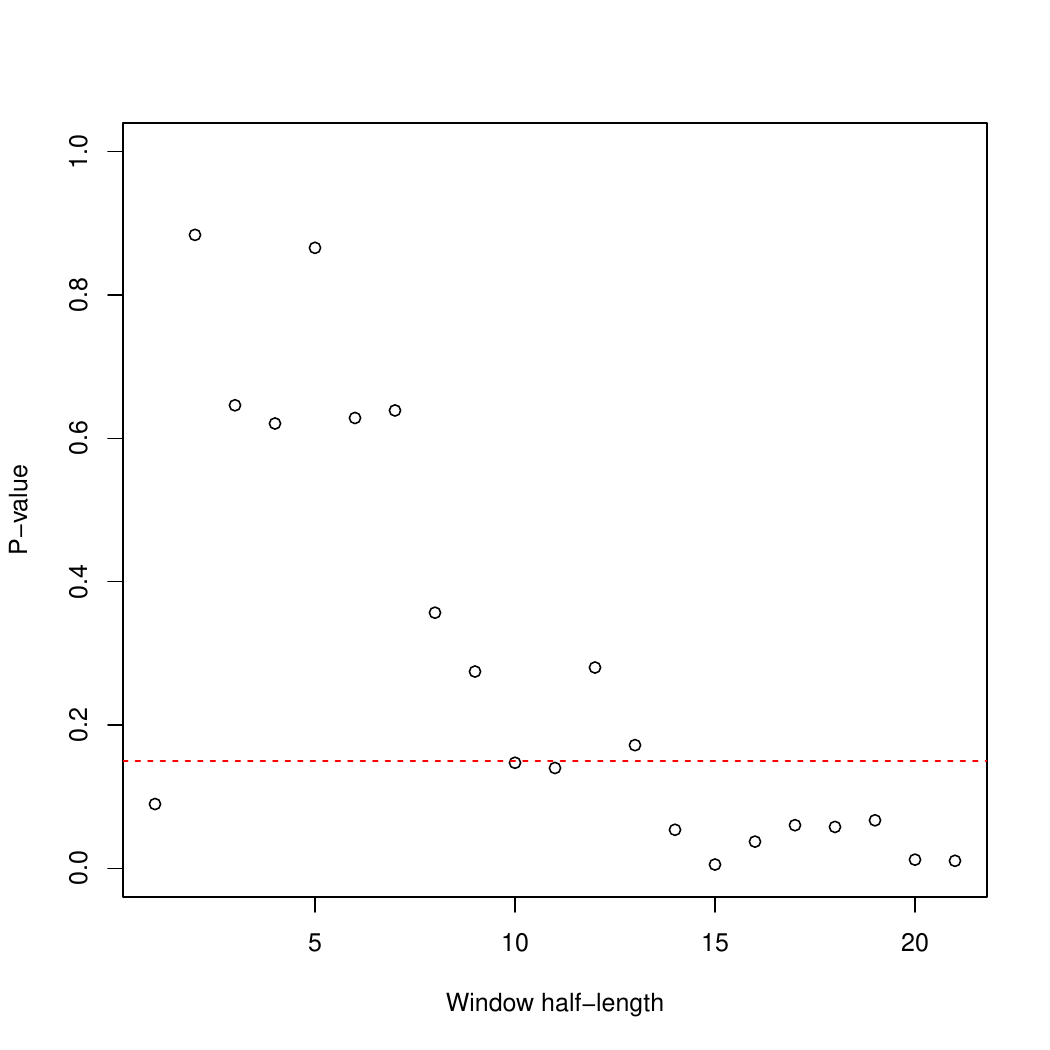} 
\caption{21 days before adoption} \label{C21}
\end{subfigure}
\hfill
\begin{subfigure}{.48\textwidth}
\centering
\includegraphics[scale=0.45]{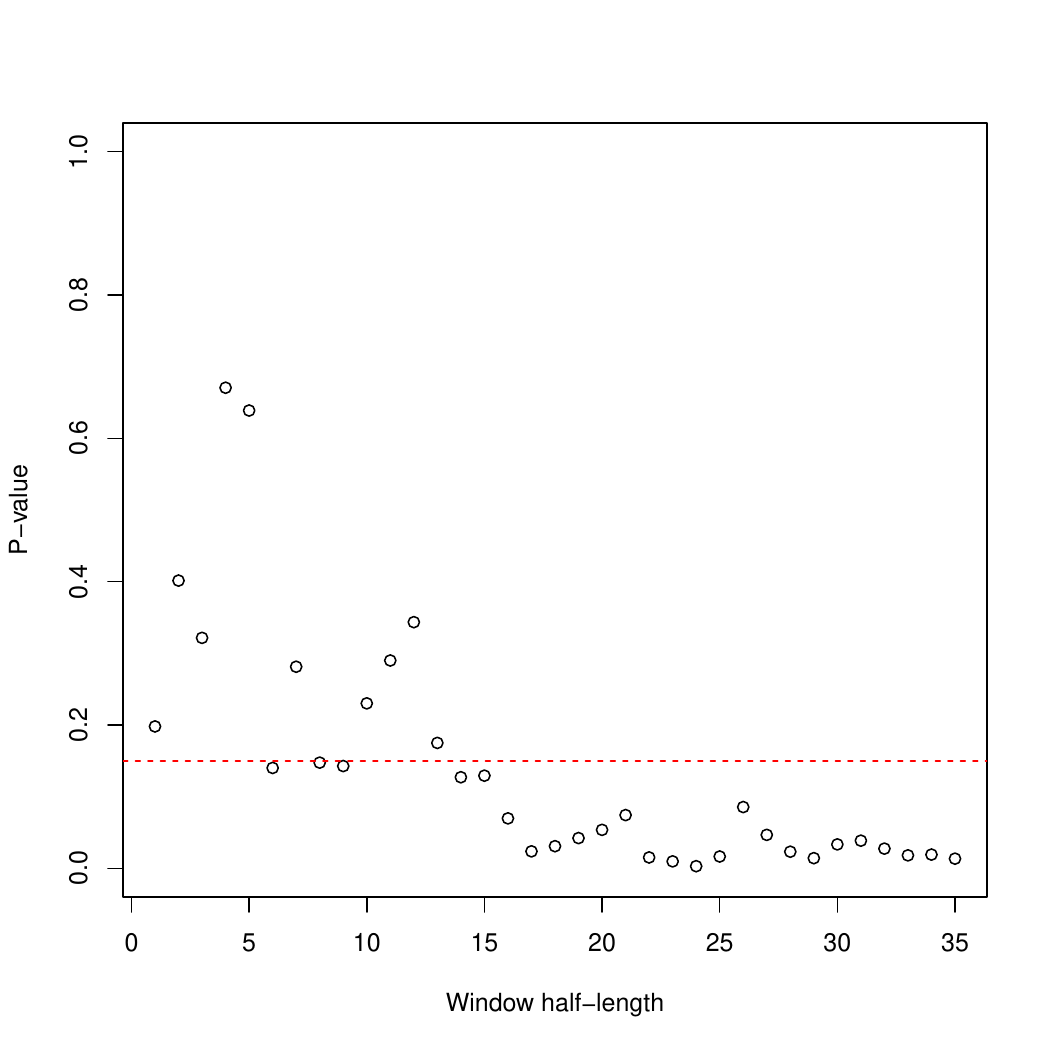} 
\caption{35 days before adoption}\label{C35}
\end{subfigure}
\caption{Window Selector Around {\it Placebo} Adoption Times}
\end{figure}

\newpage
\section{Additional Falsification Results}
\label{SAsec:falsification}

We report additional falsification results. Table 4 in the main paper presents tests of $\mathsf{H}_{\mathtt{TR}}$ and $\mathsf{H}_{\mathtt{AT}}$ with the adoption time artificially set to midnight on November 1st for the years 2015, 2016, and 2018. We show that these results remain robust when we consider alternative values of $\tau$. 

Figure \ref{fig:placebo_date} shows the average differences, $\widehat{\theta}_{\tau}$, and the randomization-based p-values for all values of $\tau$ between 1 and 21 (not just 1, 7, and 14, as reported in Table 4 in the main paper) around the different placebo adoption times  considered. When we use the actual adoption time, the null hypothesis rejected at 5\% level (grey bars are p-values) or below for almost all values of $\tau$ greater than 5, and the average differences between treated and untreated observed outcomes (red dots) are always positive and well above the values recorded for the same period in each of the placebo adoption times. In contrast, in the analyses that use placebo adoption times, p-values are generally above 10\% and, in several cases, the effects are negative (note that the red dots show only positive estimates; negative estimates would be below the horizontal axis and are omitted).  

We repeat these analyses, but instead of setting the artificial adoption time to November 1st, we set it to the same day of the week in different years. Considering that November 1st, 2017, was a Wednesday, we set the artificial adoption times to the first Wednesday of November in 2015, 2016, and 2018. The results are reported in Table \ref{tab:total-placday}, which is analogous to Table 4 in the main paper. Similarly to Table 4 in the main paper, the tests in Table \ref{tab:total-placday} fail to reject the null hypotheses of no effects. Figure \ref{fig:placebo_day} is analogous to Figure \ref{fig:placebo_date}, showing falsification results for all values of $\tau \geq 21$. 

\vspace{10pt}
\begin{table}[htbp]
\caption{Short-term Effects of CCP Reform in Montevideo\\ Around {\it Placebo} Adoption Times (Day)\\
\small{{\it Outcome}: daily number of crimes reported to police} \\
\small{{\it  Actual adoption time}:  November 1st, 2017}}
\label{tab:total-placday}\centering
\par
{ \footnotesize
\begin{tabular}{crrrrrrrrr}
\toprule
 &\multicolumn{3}{c}{2015} &\multicolumn{3}{c}{2016}&\multicolumn{3}{c}{2018}\\
\cmidrule(lr){2-4}\cmidrule(lr){5-7}\cmidrule(lr){8-10}
  &&\multicolumn{2}{c}{P-Value} &&\multicolumn{2}{c}{P-Value}&&\multicolumn{2}{c}{P-Value}\\
\cmidrule(lr){3-4}\cmidrule(lr){6-7}\cmidrule(lr){9-10}
$\tau$&\multicolumn{1}{c}{$\widehat{\theta}_{\tau}$ } &\multicolumn{1}{c}{TR}&\multicolumn{1}{c}{AT} &\multicolumn{1}{c}{$\widehat{\theta}_{\tau}$ } &\multicolumn{1}{c}{TR}&\multicolumn{1}{c}{AT} 
     &\multicolumn{1}{c}{$\widehat{\theta}_{\tau}$ } &\multicolumn{1}{c}{TR}&\multicolumn{1}{c}{AT}\\
\cmidrule(lr){1-1}\cmidrule(lr){2-4}\cmidrule(lr){5-7}\cmidrule(lr){8-10}
1 day & -0.194 & 0.689 & \multicolumn{1}{c}{-} & -0.065 & 0.910 & \multicolumn{1}{c}{-}  & 0.129 & 0.791 & \multicolumn{1}{c}{-} \\ 
  7 days & -0.106 & 0.515 & 0.513 & 0.046 & 0.772 & 0.918 & -0.081 & 0.680 & 0.864 \\ 
  14 days & -0.015 & 0.914 & 0.879 & 0.142 & 0.230 & 0.809 & 0.046 & 0.714 & 0.744 \\ 
\bottomrule
\addlinespace
 \multicolumn{10}{p{0.75\textwidth}}{ \scriptsize{Sample is 62 neighborhoods in Montevideo, each observed before and after the adoption of the CCP reform. The p-values are randomization-based for $\widehat{\theta}_{\tau} = \frac{1}{n} \sum_{i=1}^{n} \bar{Y}_{i,\tau,1} - \frac{1}{n} \sum_{i=1}^{n} \bar{Y}_{i,\tau,0}$ with 10,000 simulations. $\tau$ denotes the half-length of a symmetric window around adoption time. TR and AT refer to Treatment Reversal and Adoption Timing mechanisms, respectively. In AT mechanism, $\mathcal{A} =\{-6,-5,\ldots,0\}$ for $\tau=7$ and $\mathcal{A} =\{-13,-12,\ldots,0\}$ for $\tau=14$}.}
\end{tabular}
}
\end{table}

We also report, in Figure \ref{fig:additionalTable4} the results from Table 4 for the first 21 days of November in the years 2017, 2018, and 2019. As shown, while there is an increase on the first day of the month in both 2018 and 2019, 2017 remains the only year in which effects emerge and persist during the early days of November, under both the TR and AT assignment mechanisms. In fact, 2017 is the only case that consistently exhibits non-negative effects across all window lengths—from 1 to 21 days. 

 \begin{figure}[ht]
 \vspace{10pt}
 \centering
   \includegraphics[scale=0.35]{\inputs/CDT_Figure_SA_DailyChanges.png} 
   \caption{Daily Change at Adoption Time vs. Prior-Year Daily Changes\\ {\small (1-, 10-, and 20-Day Windows)}}
  \label{fig:dailychanges}
   \end{figure}

Aditionally, Figure \ref{fig:dailychanges} shows that the change on November 1, 2017, is exceptionally large relative to the distribution of daily changes across all 1-, 10-, and 20-day windows in the year preceding the introduction of the new CCP (November 2, 2016, to November 1, 2017).

\newpage
\section{Additional Results Assessing Potential Mechanisms}
\label{SAsec:mechanisms}

\begin{figure}[ht]
 \centering
   \includegraphics[scale=0.20]{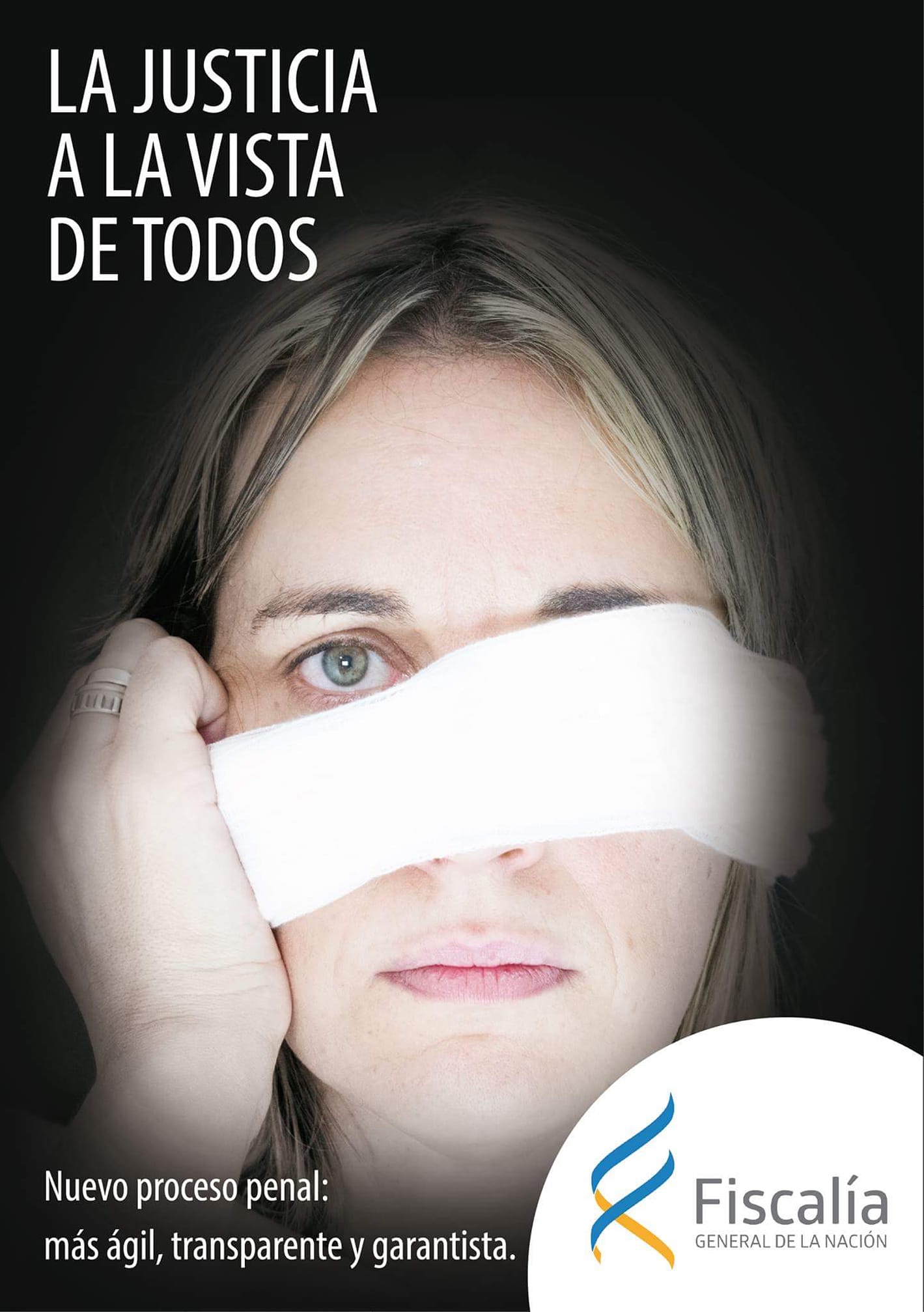}
   \caption{Billboard Promoting CCP Reform}
  \label{fig:billboard}
   \end{figure}

As we explained in the main paper, the reform was widely known to the public. The Office of the Attorney General launched a public information campaign informing the public about the reform. As a result, the reform received considerable media coverage. Figure \ref{fig:billboard} shows a billboard placed at several Montevideo bus stops highlighting the transparency and guarantees of the new procedural law.

As argued in Section 4.2 in the main paper, we have found no evidence that the increases in crime reports that we see were caused by changes in how crimes are reported rather actual crimes.  

The reform may have changed the costs victims face when reporting crimes. Although the new CCP allows individuals to report offenses directly at a prosecutor's office, our data show that nearly all crimes continued to be reported to the national police. Nonetheless, the reform may have coincided with a shift in police behavior \citep{Hausman_Kronick} that could help explain our findings. By the time the reform took effect, officers were already equipped with tablets that enabled them to file crime reports at the scene, eliminating the need for victims to travel to a police station. If the observed increase in reports was driven by enhanced street-level policing, we would expect a concentration of reports shortly after the incidents occurred. As shown in Figure \ref{fig:delay}, we find no evidence that the effects we see are driven by reports filed within 15, 30, or 45 minutes of the event. This makes it unlikely that improved on-the-spot reporting alone can account for the increase we document.

We also compare the immediate impact of the new CCP on the number of reported thefts to its impact on the number of reported robberies (i.e., violent thefts) and domestic violence (the most frequent crime against persons). If the increase in crime reports is due to a more selective use of preventive detention, we should not observe any impact on domestic violence (a crime that is considered very severe by prosecutors under both procedural regimes), while the effects should be present for thefts and robberies---and should be higher for thefts than for robberies, as thefts are classified as non-violent and thus might avoid preventive detention.

Table~\ref{tab:types}  presents results. As expected, the number of domestic violence incidents reported to the police does not exhibit an immediate response to the change in CCP. In fact, negative effects are observed for $\tau=1$ and $\tau=7$ (null hypothesis not rejected). The effects for robberies also appear weak, with a decrease in the smallest window. Thefts is the only category that shows a consistent increase in all windows considered, although the p-values are higher than for the total effect, particularly for the AT mechanism.

This pattern suggests that the increase in the total number of police reports is not driven by the two most frequent violent crimes. In contrast, the results show that the reform might have increased thefts immediately after implementation. However, these effects are not as strong as the effects for total crime reported in Table 2, which suggests that the decrease in preventive detention is likely only part of the explanation.

\vspace{10pt}
\begin{table}[htbp]
\caption{Short-Term Effects of CCP Reform in Montevideo\\ for Different Types of Crime\\
\small{{\it Outcome}: daily number of crimes reported to police} \\
\small{{\it  Actual adoption time}: November 1st, 2017}}
\label{tab:types}\centering
\par
{ \footnotesize
\begin{tabular}{crcccccc}
\toprule
\multicolumn{7}{c}{Property Crime: {\it Theft}}\\
\midrule 
 &\multicolumn{2}{c}{Estimates} &\multicolumn{2}{c}{TR Mechanism}&\multicolumn{2}{c}{AT Mechanism} \\
\cmidrule(lr){2-3}\cmidrule(lr){4-5}\cmidrule(lr){6-7}
$\tau$&\multicolumn{1}{c}{$\widehat{\theta}_{\tau}$}  &\multicolumn{1}{c}{$\frac{1}{n} \sum_{i=1}^{n} \bar{Y}_{i,\tau,0}$} & p-value & 95\% CI & p-value & 95\% CI \\
\cmidrule(lr){1-1}\cmidrule(lr){2-2}\cmidrule(lr){3-3}\cmidrule(lr){4-5}\cmidrule(lr){6-7}

1 day & 0.500 & 2.290 & 0.115 & [-0.09, 1.10] &  - & - \\ 
7 days & 0.194 & 2.270 & 0.067 & [ 0.00, 0.39] & 0.338 & [-0.40, 0.71] \\ 
14 days & 0.150 & 2.363 & 0.162 & [-0.05, 0.35] & 0.185 & [-0.04, 1.27] \\ 
\midrule
\multicolumn{7}{c}{Violent Crime: {\it Robbery}}\\
\midrule 
 &\multicolumn{2}{c}{Estimates} &\multicolumn{2}{c}{TR Mechanism}&\multicolumn{2}{c}{AT Mechanism} \\
\cmidrule(lr){2-3}\cmidrule(lr){4-5}\cmidrule(lr){6-7}
$\tau$&\multicolumn{1}{c}{$\widehat{\theta}_{\tau}$}  &\multicolumn{1}{c}{$\frac{1}{n} \sum_{i=1}^{n} \bar{Y}_{i,\tau,0}$} & p-value & 95\% CI & p-value & 95\% CI \\
\cmidrule(lr){1-1}\cmidrule(lr){2-2}\cmidrule(lr){3-3}\cmidrule(lr){4-5}\cmidrule(lr){6-7}
1 day & -0.258 & 0.935 & 0.207 & [-0.63, 0.11] & - & - \\ 
7 days & 0.097 & 0.707 & 0.088 & [-0.01, 0.20] & 0.700 & [-0.35, 0.19] \\ 
14 days & 0.021 & 0.719 & 0.676 & [-0.07, 0.11] & 0.650 & [-0.20, 0.22] \\ 
\midrule 
\multicolumn{7}{c}{Violent Crime: {\it Domestic Violence}}\\
\midrule 
 &\multicolumn{2}{c}{Estimates} &\multicolumn{2}{c}{TR Mechanism}&\multicolumn{2}{c}{AT Mechanism} \\
\cmidrule(lr){2-3}\cmidrule(lr){4-5}\cmidrule(lr){6-7}
$\tau$ &\multicolumn{1}{c}{$\widehat{\theta}_{\tau}$}  &\multicolumn{1}{c}{$\frac{1}{n} \sum_{i=1}^{n} \bar{Y}_{i,\tau,0}$} & p-value & 95\% CI & p-value & 95\% CI \\
\cmidrule(lr){1-1}\cmidrule(lr){2-2}\cmidrule(lr){3-3}\cmidrule(lr){4-5}\cmidrule(lr){6-7}
 1 day & -0.097 & 0.613 & 0.560 & [-0.37, 0.17] & - & - \\ 
 7 days & -0.035 & 0.558 & 0.494 & [-0.12, 0.05] & 0.420 & [-0.35, 0.14] \\ 
 14 days & 0.020 & 0.533 & 0.571 & [-0.04, 0.08] & 0.751 & [-0.24, 0.19] \\ 
\bottomrule
\addlinespace
 \multicolumn{7}{p{0.75\textwidth}}{ \scriptsize{Sample is 62 neighborhoods in Montevideo, each observed before and after the adoption of the CCP reform. The p-values are randomization-based for $\widehat{\theta}_{\tau} = \frac{1}{n} \sum_{i=1}^{n} \bar{Y}_{i,\tau,1} - \frac{1}{n} \sum_{i=1}^{n} \bar{Y}_{i,\tau,0}$ with 10,000 simulations. $\tau$ denotes the half-length of a symmetric window around adoption time. CI refers to confidence interval, calculated by inversion based on constant treatment effect model. TR and AT refer to Treatment Reversal and Adoption Timing mechanisms, respectively. In AT mechanism, $\mathcal{A} =\{-6,-5,\ldots,0\}$ for $\tau=7$ and $\mathcal{A} =\{-13,-12,\ldots,0\}$ for $\tau=14$}.}
\end{tabular}
}
\end{table}

\newpage
\section{Results Using Event Study Methods}
\label{SAsec:eventstudy}

We estimate effects using event study methods based on linear models, using $i=1,\ldots n$ units and $t=1,\ldots,T$ time periods. For implementation, we use the approach by \cite{FreyaldenhovenEtal2019-AER,FreyaldenhovenEtal2024-wp}, and fit a linear model of the outcome ($q_{it}$) on a unit fixed effect ($\alpha_i$), a time fixed effect ($\gamma_t$), and binary policy variables that indicate whether each unit is treated in specific periods ($\sum_{j=-G}^{M} \beta_j z_{i,t-j}$), where the parameters $\beta_j$ capture the effects of the policy. We created event study plots based on this fit, using the \texttt{xtevent} command \citep{FreyaldenhovenEtal2025-Stata}. 

\begin{figure}[h]
\vspace{-0.1in}
\caption{Event Study Plots for CCP Effect}
\label{fig:ESplots}
\centering
    \begin{subfigure}{.48\textwidth}
    \centering
    \includegraphics[scale=0.4]{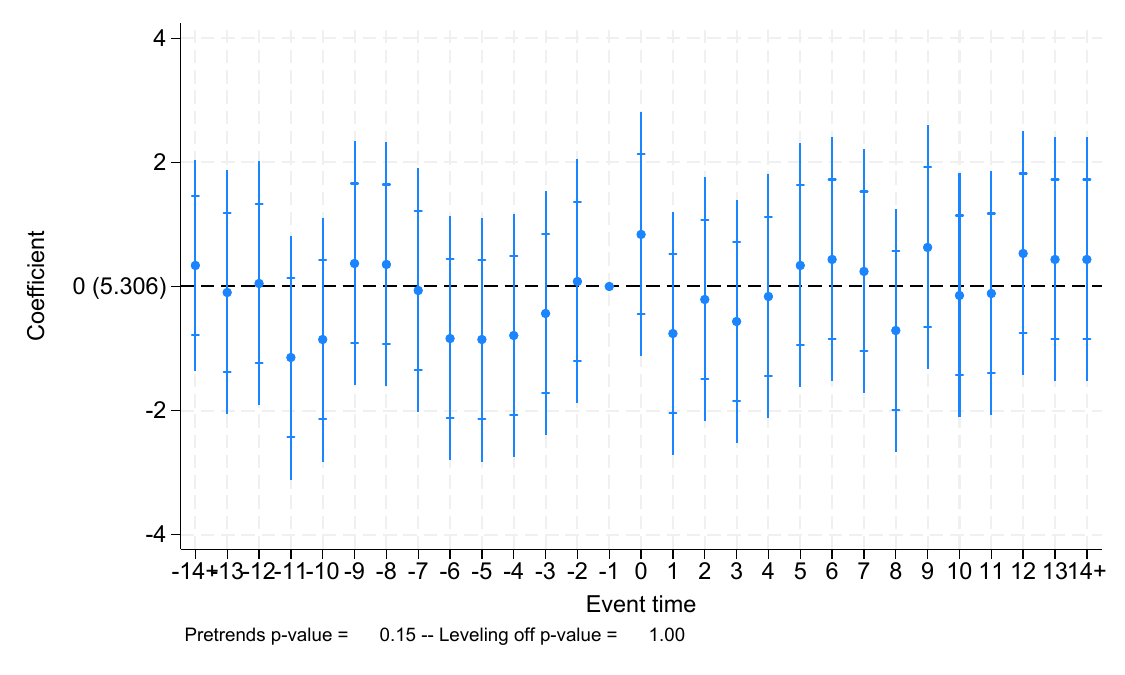} 
    \caption{$\pm$ 15 days, no fixed/time effects} \label{ES1}
    \end{subfigure}
    \hfill
    \begin{subfigure}{.48\textwidth}
    \centering
    \includegraphics[scale=0.4]{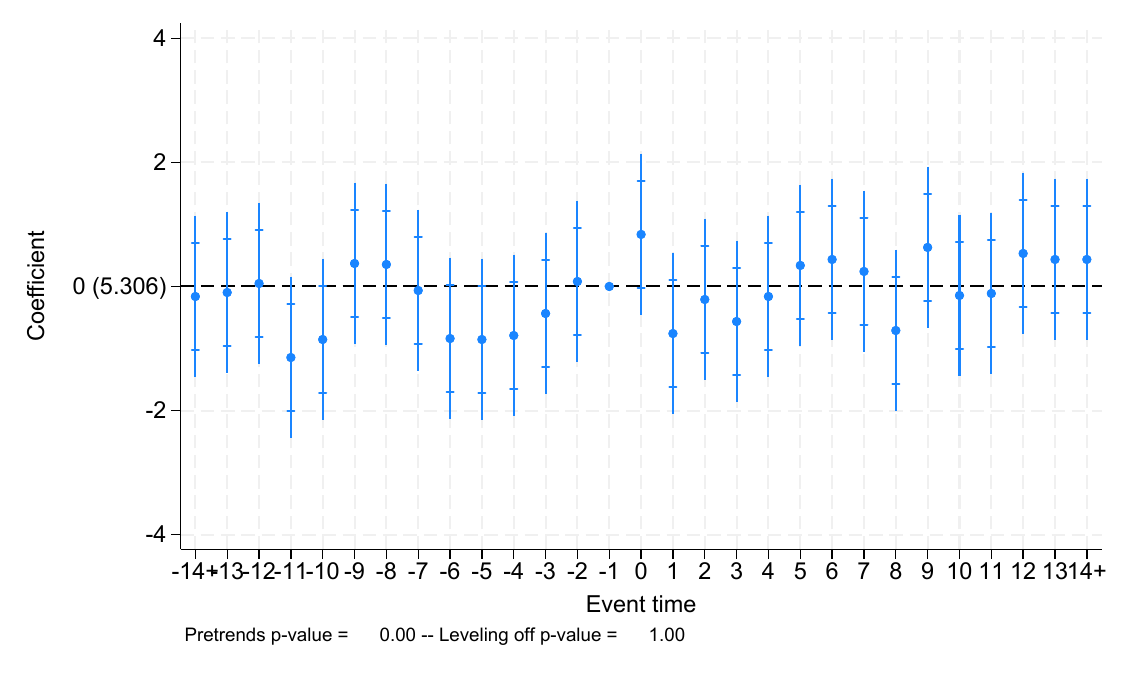} 
    \caption{$\pm$ 15 days, with fixed/time effects} \label{ES2}
    \end{subfigure}
    \begin{subfigure}{.48\textwidth}
    \centering
    \includegraphics[scale=0.4]{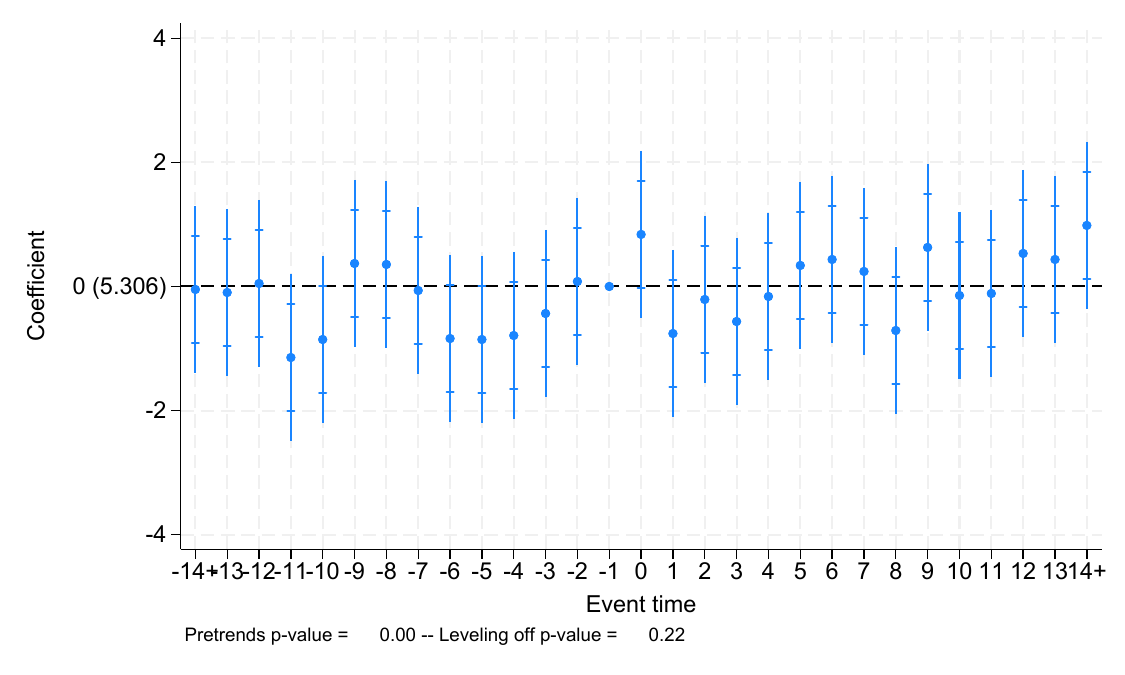} 
    \caption{$\pm$ 50 days, with fixed/time effects} \label{ES3}
    \end{subfigure}
    \hfill
    \begin{subfigure}{.48\textwidth}
    \centering
    \includegraphics[scale=0.4]{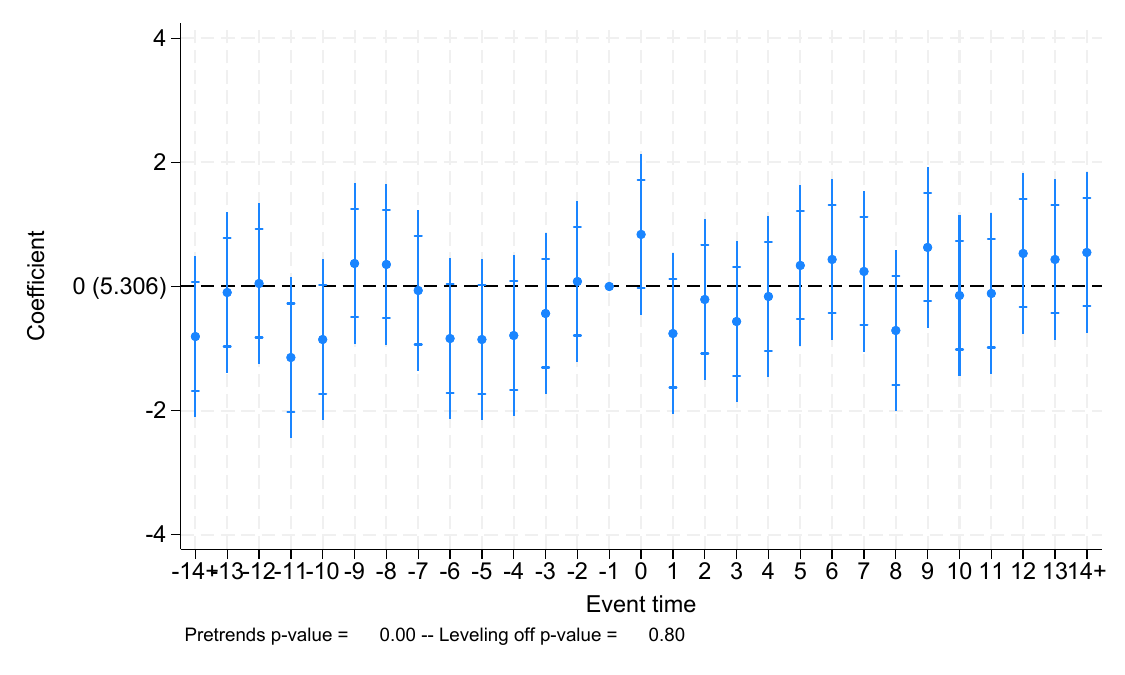} 
    \caption{$\pm$ 100 days, with fixed/time effects} \label{ES4}
    \end{subfigure}
\end{figure}

We fit this model in four different ways: (a) using 15 days before and 15 days after adoption, and excluding fixed-effects and time-effects; (b) using 15 days before and 15 days after adoption, including fixed-effects and time-effects; (c) using 50 days before and 50 days after adoption, including fixed-effects and time-effects; and  (d) using 100 days before and 100 days after adoption, including fixed-effects and time-effects. The event-study plots are shown in Figure \ref{fig:ESplots}, where the outcome at the time period immediately before adoption is normalized to zero.

In Figure \ref{fig:ESplots}(\subref{ES1}), we exclude time effects and fixed effects, as both types of effects would be estimated with relatively few observations (only 30 periods for fixed effects, only 62 units for time effects). Overall, the cumulative effects show null effects between day -14 and day 14 after adoption of the intervention. Judging from these results, one would conclude that the short-run effects of the policy are zero. Adding fixed effects and time effects and keeping the number of days at 15, as shown in Figure \ref{fig:ESplots}(\subref{ES2}), or adding additional days, as shown in Figure \ref{fig:ESplots}(\subref{ES3}) and \ref{fig:ESplots}(\subref{ES4}), does not change the overall pattern. 

Moreover, for each model, we conducted a test that the coefficients capturing the cumulative effect of the policy at time 0, 7, and 14 are jointly zero using an F-test. Table \ref{tab:ESFtests} shows that the results vary considerably depending on how the model is specified. When only 15 days are used before/after adoption, the null hypothesis that the effects are jointly zero at time 0, 7 and 14 cannot be rejected, with very high p-value above 0.60. The same conclusion is drawn from model (b), where fixed effects and time effects are included. However, when the number of periods used to fit the linear model is increased to 50 days before/after adoption in model (c), the p-value drops to below 0.08. The p-value increases again for model (d). This illustrates how results can be sensitive to the parametric specification that is used to fit the model.

\begin{table}
\caption{Test that Cumulative CCP Effect at Time 0, 7 and 14 Are Jointly Zero}
\label{tab:ESFtests}
\begin{tabular}{ccc}
Model & F &  $\mathtt{Prob} > F$ \\
\midrule
(a)       & 0.58 &   0.6268\\ 
(b)      & 1.29& 0.2766\\
(c)    &  2.26 &  0.0792\\
 (d)    & 1.35 & 0.2566\\
\bottomrule \\    
\end{tabular}
\end{table}

\clearpage
\onehalfspacing
\bibliographystyle{jasa}
\bibliography{CDT_2025_JRSSA--bib}

\begin{landscape}
\begin{figure}[!ht]
\includegraphics[width=\textwidth]{\inputs/CDT_Figure_SA_PlaceboDate.png} 
\caption{Effects of CCP Reform for Different Windows Around {\it Placebo} Adoption Times ({\it Date})}
\label{fig:placebo_date}
\end{figure}
\end{landscape} 

\begin{landscape}
    \begin{figure}[ht]
    \includegraphics[width=\textwidth]{\inputs/CDT_Figure_SA_PlaceboDay.png} 
    \caption{Effects of CCP Reform for Different Windows Around {\it Placebo} Adoption Times ({\it Day})}
    \label{fig:placebo_day}
    \end{figure}
\end{landscape}

\begin{landscape}
\begin{figure}[!ht]
\includegraphics[width=\textwidth]{\inputs/CDT_Figure_SA_ReportingDelay.png} 
\caption{Effects of CCP Reform When Restricting {\it Reporting Delay}}
\label{fig:delay}
\end{figure}
\end{landscape}

\begin{landscape}
    \begin{figure}[ht]
    \includegraphics[width=\textwidth]{\inputs/CDT_Figure_SA_PlaceboDate1819.png} 
    \caption{Effects of CCP Reform for Different Windows Around {\it Placebo} Adoption Times, Additional Years ({\it Date})}
    \label{fig:additionalTable4}
    \end{figure}
\end{landscape}